\documentstyle[epsfig,bm,amsmath,amssymb,aps,12pt,tighten]{revtex}

\allowdisplaybreaks[3]

\def\Fs#1{{}\kern-.45em \not \kern-.12em #1\hspace{.2pt}}

\def\as{\alpha_s}
\def\A{{\cal A}}
\def\B{{\cal B}}

\def\M{{\cal M}}
\def\I{{\cal I}}
\def\O{{\cal O}}
\def\gs{{\gamma^*}}

\newcommand{\cut}{\alpha_\ell^{\mathrm cut}}
\newcommand{\ellq}{\bm\ell''}
\newcommand{\K}{{\cal K_{\mathrm real}}}
\newcommand{\mscr}[1]{\mbox{\scriptsize{$#1$}}}
\newcommand{\amax}{\alpha_\ell^{\mathrm max}}
\newcommand\npb[3]{{Nucl. Phys. }{\bf B #1} (#2) #3}
\newcommand\plb[3]{{Phys. Lett. }{\bf B #1} (#2) #3}

\newcommand{\klr}{(\bm{k}+\bm{\ell}+\bm{r})}
\newcommand{\kl}{(\bm{k}+\bm{\ell})}
\newcommand{\kr}{(\bm{k}+\bm{r})}
\newcommand{\lmr}{(\bm{\ell}-\bm{r})}
\newcommand{\lpr}{(\bm{\ell}+\bm{r})}
\newcommand{\lmrr}{(\bm{\ell}-2\bm{r})}
\newcommand{\klkr}{(\bm{k}+\bm{\ell})(\bm{k}+\bm{r})}
\newcommand{\lkl}{\bm{\ell}(\bm{k}+\bm{\ell})}
\newcommand{\rkl}{\bm{r}(\bm{k}+\bm{\ell})}
\newcommand{\lkr}{\bm{\ell}(\bm{k}+\bm{r})}
\newcommand{\rkr}{\bm{r}(\bm{k}+\bm{r})}

\newcommand{\kkl}{\bm{k}(\bm{k}+\bm{\ell})}
\newcommand{\kkr}{\bm{k}(\bm{k}+\bm{r})}
\newcommand{\kklr}{\bm{k}(\bm{k}+\bm{\ell}+\bm{r})}

\newcommand{\rklr}{\bm{r}(\bm{k}+\bm{\ell}+\bm{r})}
\newcommand{\krklr}{(\bm{k}+\bm{r})(\bm{k}+\bm{\ell}+\bm{r})}
\newcommand{\klklr}{(\bm{k}+\bm{\ell})(\bm{k}+\bm{\ell}+\bm{r})}
\newcommand{\ep}{\epsilon}

\begin{document}

\tighten

\title{
\hfill\parbox{10cm}{\normalsize\raggedleft Cavendish-HEP-02/04\\
DESY 02-114\\
hep-ph/0208130}\\[30pt]
NLO Corrections to the Photon Impact Factor:\\
Combining Real and Virtual Corrections}

\author{J.~Bartels$^{(a)}$\thanks{supported by the EU
  TMR-Network `QCD and the Deep Structure of Elementary Particles',
  contract number FMRX-CT98-0194 (DG 12-MIHT).}, 
D.~Colferai$^{(a)*}$\thanks{Alexander von Humboldt Fellow.}, 
S.~Gieseke$^{(b)*}$, 
A.~Kyrieleis$^{(a)*}$\thanks{supported by the Graduiertenkolleg
  `Zuk\"unftige Entwicklungen der Teilchenphysik'.}\\[10pt]}
\bigskip

\address{(a) II.~Institut f\"ur Theoretische Physik, Universit\"at
  Hamburg, \\
  Luruper Chaussee 149, 22761 Hamburg, Germany\\
  (b) Cavendish Laboratory, University of Cambridge, \\
  Madingley Road, Cambridge CB3 0HE, U.K.}

\maketitle

\bigskip
\begin{abstract}

  \noindent
  In this third part of our calculation of the QCD NLO corrections to
  the photon impact factor we combine our previous results for the
  real corrections with the singular pieces of the virtual corrections
  and present finite analytic expressions for the
  quark-antiquark-gluon intermediate state inside the photon impact
  factor. We begin with a list of the infrared singular pieces of the
  virtual correction, obtained in the first step of our program.  We
  then list the complete results for the real corrections
  (longitudinal and transverse photon polarization). In the next step
  we define, for the real corrections, the collinear and soft singular
  regions and calculate their contributions to the impact factor.
  We then subtract the contribution due to the central region.
  Finally, we combine the real corrections with the singular pieces of the 
  virtual corrections and obtain our finite results.
\end{abstract}

\section{Introduction}
\label{sec:introduction}

\noindent
This paper represents the third part of our investigation of the photon impact
factor in next-to-leading order QCD.  First results of the calculation of NLO
corrections to the photon impact factor have been published in \cite{BGQ} and
\cite{BGK}. In \cite{BGQ} we have computed the QCD NLO corrections to the
process $\gamma^*+q \to (q\bar{q})+q$ which lead to the virtual corrections of
the photon impact factor. The second paper, \cite{BGK}, contains the process
$\gamma^*+q\to (q\bar{q}g) + q$, leading to the real corrections of the photon
impact factor.  Results are given for the helicity-summed squared matrix
elements, with the longitudinally polarized photon in the initial state.  In the
present paper we will complete this part of our calculation by listing also the
results for the transverse polarization of the photon.

The main purpose of the present paper is the combination of singular pieces of
virtual and real corrections. To begin with the former ones, in \cite{BGQ} we
have listed the results for the virtual corrections which contain both the
finite and the singular pieces. As the first step we therefore have to separate
the sum of all singular $1/\epsilon$ terms from the finite part.  In the next
step we turn to the real corrections. Infrared singularities will come from
those kinematic regions of the $q\bar{q}g$ system where the gluon is either soft
or collinear with the quark or with the antiquark. We compute the soft and the
collinear approximations of our matrix elements, and, by subtracting these
divergent contributions from the full matrix elements, we form finite
combinations. Finally, by combining the singular pieces of the real corrections
with those coming from the virtual corrections, we obtain further finite terms
which have the form of the Born approximation and can be computed analytically.
As the main result of this paper, we present finite expressions for the
$q\bar{q}g$ contributions to the photon impact factor. As the fourth and
remaining part of our program we will be left with the task of evaluating
numerically the finite integrals obtained in this paper, as well as the finite
pieces from the virtual corrections.

This paper will be organized as follows. After the definition of the impact
factor (section~\ref{sec:definition}) we first
(section~\ref{sec:singular-virtual}) list the infrared singular pieces of the
virtual corrections. We then turn to the real corrections and present the
complete results for the real corrections, both for the longitudinal and for the
transverse photon (section~\ref{sec:real}).  In the following
section~\ref{sec:singular-real} we compute the collinear and the soft limits of
the real corrections.
We discuss in detail the definition of the impact factor correction in
section~\ref{sec:NLOIF}, where the subtraction of the central region, i.e.
the leading $\log s$ term, is discussed.
In section~\ref{sec:finite} we complete
our program, by combining the singular pieces of the real corrections
with those of the virtual corrections and by demonstrating their infrared
finiteness. 
In a final section we summarize our results and give a brief outline
of the remaining part of our program.

\section{Definition of the Impact Factor}
\label{sec:definition}

\noindent
We consider the elastic scattering of two particles $A$ and $B$ with momenta
$p_A$ and $p_B$ in the Regge limit $s \to \infty$, where the momentum transfer
$t=q^2$ is kept fixed. In leading order in $\alpha_s$ we write the scattering
amplitude $T_{AB}$ in the form:
\begin{equation}
  T_{AB}^{(0)} (s, t = q^2) = is \int \frac{d^{D-2}\bm r}{(2\pi)^{D-2}}
  \tilde\Phi_{A}^{(0)} \frac{1}{\bm r^2}\frac{1}{(\bm q-\bm r)^2} 
\tilde\Phi_{B}^{(0)}\,. 
\end{equation}
The impact factors $\Phi_A^{(0)}$ and $\Phi_B^{(0)}$ for the external particles
(quarks, gluons, or photons) are of order $\alpha_s$. The exchanged gluons carry
transverse momenta $\bm r$ and $(\bm q-\bm r)$ respectively.  Taking the cut
amplitude we define the total cross section for the scattering of $A$ and $B$
via the optical theorem as
\begin{equation}
  \label{eq:factorized}
  \sigma_{AB}^{(0)} = \frac{1}{s} {\mathrm{Im}} T_{AB}^{(0)} (s, t=0)
  = \int \frac{d^{D-2}\bm r}{(2\pi)^{D-2}}
  \Phi_{A}^{(0)} \frac{1}{\bm r^4}\Phi_{B}^{(0)}\,. 
\end{equation}
Here we have dropped the tilde symbol on the impact factors to denote
the additional $s$-channel cut. This equation can be used as the
definition of the leading order impact factors $\Phi_A^{(0)}$ and
$\Phi_B^{(0)}$. Note that we have summed over the colour indices of
the $t$-channel gluons implicitly.  The total cross section can also be
written as
\begin{equation}
  \label{eq:sigfeyn}
  \sigma_{AB}^{(0)} = \frac{1}{2s} \int |\M_{AB}^{(0)}|^2 d\phi\, , 
\end{equation}
where $1/2s$ denotes the flux factor (in the high energy limit),
$\M_{AB}^{(0)}$ the matrix element for the scattering process $A+B \to
A'+ B'$ with the exchange of a gluon, and $d\phi$ stands for the phase
space of the final state $A'+B'$. For our case of interest,
$\gamma^*q$ (or $\gamma^*\gamma^*$) scattering, $A'$ denotes a
$q\bar{q}$ pair, $B'$ a quark (or a $q\bar{q}$ pair).
\begin{figure}[htbp]
  \begin{center}
    \epsfig{file=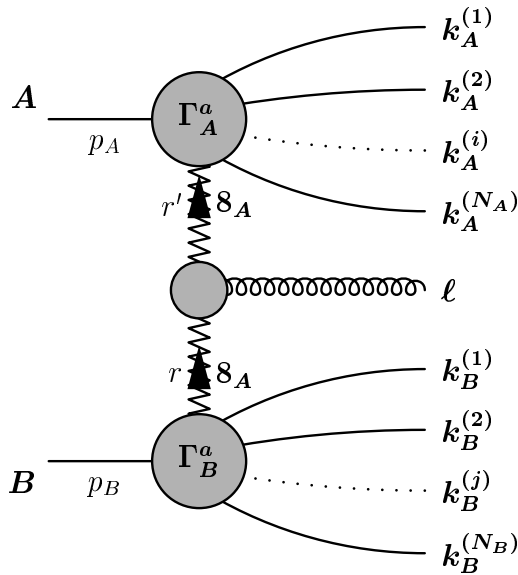} \bigskip
    \caption{Kinematics for $A+B\to A'+g+B'$ with initial state momenta
      $p_A, p_B$ and final state momenta $k_A^i, \ell, k_B^j$. An
      ${\boldmath 8_A}$-reggeon is exchanged, emitting a real gluon
      with momentum $\ell$ which is assumed to be separated from the
      other final state particles by large rapidity gaps.
      \label{fig:ifdefkin}}
  \end{center}
\end{figure}
\noindent
In the high energy limit, the scattering amplitude $\M_{AB}$ for the scattering
$A+B\to A'+B'$ (with the invariant masses of the particles or systems $A'$ and
$B'$ being finite) 
(Fig.~\ref{fig:ifdefkin} without the extra produced gluon) 
is described by the exchange of a
reggeized gluon and can be written in the following form:
\begin{equation}
\label{eq:regge}  
\M_{AB} = \frac{s}{t}\Gamma_{A\to A'}^{a}
  \left[\left(\frac{\hfill s}{-t}\right)^{\omega(t)} 
  + \left(\frac{-s}{-t}\right)^{\omega(t)} \right]\Gamma_{B \to B'}^{a} \,. 
\end{equation} 
Here, $\omega(t)$ stands for the Regge trajectory of the gluon, and
$\Gamma_{A \to A'}^{a}$ and $\Gamma_{B \to B'}^{a}$ are the
particle-particle-reggeon vertices.  The index $a$ denotes the colour
of the exchanged reggeized gluon.  Eq.(\ref{eq:regge}) exhibits the
factorization property of Regge theory: when written in complex
angular plane, the residue of the gluon Regge pole can be written as a
product of two vertex functions, one for the incoming particle $A \to
A'$, the other one for $B \to B'$. In the total cross section formula
(\ref{eq:sigfeyn}) this factorization property leads to the
factorization of impact factors. Since (\ref{eq:regge}) is valid not
only in leading order, also the validity of the impact factor
representation is rather general. However, when going beyond the
leading order, we have to replace the $t$-channel propagators of the
exchanged gluons by reggeon propagators. In leading order, we put
$\omega(t)$ in (\ref{eq:regge}) equal to zero and obtain
\begin{equation}
  \label{eq:regge0}
  \M_{AB}^{(0)} = \Gamma_{A \to A'}^{(0),a} \frac{2s}{t}
\Gamma_{B \to B'}^{(0),a}\, . 
\end{equation}
Inserting this expression into (\ref{eq:sigfeyn}), we arrive at  
(\ref{eq:factorized}).
  
In the next order, $\alpha_s^3$, eq.(\ref{eq:factorized}) receives
several corrections.  In order to illustrate the general pattern it is
instructive to start from (\ref{eq:sigfeyn}). There are three new
contributions on the right-hand-side of (\ref{eq:sigfeyn}). The first
one comes from the same intermediate state $A'$, $B'$ as in the
leading order case, but there are higher order corrections inside the
vertex functions $\Gamma_{A \to A'}$ or $\Gamma_{B \to B'}$ of the
scattering amplitude $A+B \to A'+B'$ (or its complex conjugate). A
second correction is due to the reggeization of the exchanged gluon:
from (\ref{eq:regge}) it follows that in next-to-leading order
reggeization provides an extra term in $\M_{AB}$ proportional to
$\Gamma_{A \to A'}^{(0),a} \omega(t) \ln (s/-t) \Gamma_{B \to
  B'}^{(0),a}$. These two corrections will be referred to as `virtual'
corrections to the impact factor.  The third correction comes from the
production of an extra gluon in the intermediate state (`real'
corrections): $A+B \to A'+g+B'$ (Fig.~\ref{fig:ifdefkin}). 
In the high energy limit, we divide
this intermediate state into configurations with one or two large
rapidity gaps: in the former case, the extra gluon belongs to the
fragmentation region of $A$ or $B$, in the latter case to the central
(or: multiperipheral) region.

Correspondingly, we separate the integral over the phase space $d\phi$.
Begin with the configuration with one single rapidity gap.  If
$N_{A'}$ ($N_{B'}$) particles in the final state belong to the system
$A'$ ($B'$) and carry momenta $k_{A'}^i$ ($k_{B'}^j$), we may write
the phase space in $D$ dimensions explicitly as
\begin{equation}
  d\phi = d\tilde\phi_{A'} d \tilde\phi_{B'}
  (2\pi)^D 
  \delta^{(D)}(p_A+p_B - {\textstyle\sum_i} k_{A'}^i - {\textstyle\sum_j} 
k_{B'}^j)
\end{equation}
with
\begin{equation}
  d\tilde\phi_{A'} = 
  \prod_{i=1}^{N_A} 
  \frac{d^D k_{A'}^i}{(2\pi)^{D-1}} \delta^+(k_{A'}^i{}^2 - m_A^i{^2})  
\end{equation}
and
\begin{equation}
  d\tilde\phi_{B'} = 
  \prod_{j=1}^{N_B} 
  \frac{d^D k_{B'}^j}{(2\pi)^{D-1}} \delta^+(k_{B'}^j{}^2 - m_B^j{^2})  \,.
\end{equation}
Introducing the explicit integration over the reggeon momentum $r$ with a
Sudakov decomposition $r = \alpha_r q'+ \beta_r p + r_\perp$ via one extra delta
function and
\begin{equation}
  \frac{d^D r}{(2\pi)^D} = \frac{s}{2} \frac{d\alpha_r}{2\pi}
  \frac{d\beta_r}{2\pi} \frac{d^{D-2}\bm r}{(2\pi)^{D-2}}
\end{equation}
we may write
\begin{equation}
  \label{eq:ps2}
  d\phi = 
  d\phi_{A'} \frac{s d\beta_r}{2\pi}
  d\phi_{B'} \frac{s d\alpha_r}{2\pi}
  \frac{1}{2s}
  \frac{d^{D-2}\bm r}{(2\pi)^{D-2}}\,,
\end{equation}
where $d\phi_{A'} = d\tilde\phi_{A'} (2\pi)^D \delta^{(D)}(p_A+
r-{\textstyle\sum_i} k_{A'}^i)$ and $d\phi_{B'} = (2\pi)^D
d\tilde\phi_{B'}\delta^{(D)}(p_B- r- {\textstyle\sum_j} k_{B'}^j)$ are
the usual phase space measures for the final state particles belonging
to the systems $A'$ and $B'$ respectively. For an intermediate state
with two rapidity gaps (gluon in the central region) we have two
$t$-channel reggeon momenta, $r$ and $r'$. Denoting the momentum of
the produced gluon by $\ell\equiv r-r'=\alpha_\ell q' + \beta_\ell p
+\ell_{\perp}$ eq.~(\ref{eq:ps2}) will be generalized to:
\begin{equation}
\label{eq:ps3}
d\phi = (\frac{1}{2s})^2\;\;
        d\phi_{A'} \frac{s d\beta_r}{2\pi}\;\;
        \frac{s d\alpha_r}{2\pi} d\phi_g \frac{s d\beta_{r'}}{2\pi}\;\;
        \frac{s d\alpha_{r'}}{2\pi} d\phi_{B'}\;\;
        \frac{d^{D-2}\bm r}{(2\pi)^{D-2}}\frac{d^{D-2}\bm r'}{(2\pi)^{D-2}}
\end{equation}
where
\begin{equation}
\label{eq:psg}
d\phi_g= 
  \frac{d^D \ell}{(2\pi)^{D-1}} \delta^+(\ell^2) 
       (2\pi)^D \delta^{(D)}(r-r'+\ell)
\end{equation}
is the phase space measure for the gluon in the central region.

Returning to the impact factor representation (\ref{eq:factorized}),
the NLO corrections due to the extra gluon in the $s$-channel can be
grouped into two parts.  The gluon in the fragmentation region of $A$
or $B$ leads to corrections of the impact factor $\Phi_A$ or $\Phi_B$,
respectively, whereas the gluon in the central region represents the
first rung of the BFKL ladder (in the leading-logarithmic
approximation) \cite{BFKL}. Together with the other NLO corrections
mentioned before, the entire sum of NLO corrections to the total cross
section can be cast into the following form:
\begin{align}\nonumber
  \sigma_{AB}^{(1)} &= \frac{1}{s} {\mathrm{Im}}\, T_{AB}^{(1)} (s, t=0)\\
&= \int \frac{d^{D-2}\bm r}{(2\pi)^{D-2}}
  \Phi_{A}^{(1)} \frac{1}{\bm r^4}\Phi_{B}^{(0)} + 
   \int \frac{d^{D-2}\bm r}{(2\pi)^{D-2}}
  \Phi_{A}^{(0)} \frac{1}{\bm r^4}\Phi_{B}^{(1)} \nonumber \\
&\quad +  
     \int \frac{d^{D-2}\bm r}{(2\pi)^{D-2}}
  \Phi_{A}^{(0)} \ln (s/{\bm r}^2) 2\omega^{(1)}({\bm r}^2) 
     \frac{1}{\bm r^4}\Phi_{B}^{(0)} \nonumber \\
&\quad+ \ln (s/s_0)
\int \frac{d^{D-2}\bm r}{(2\pi)^{D-2}} \frac{d^{D-2}\bm r'}{(2\pi)^{D-2}}
  \Phi_{A}^{(0)} \frac{1}{\bm r^4} \K(\bm r, \bm r')\frac{1}{{\bm r'}^4} 
\Phi_{B}^{(0)}  \,.   \label{eq:svil} 
\end{align}
Here $\K(\bm r, \bm r')$ denotes the square of the gluon production
vertex, the BFKL kernel (sometimes referred to as the `real' part of
the BFKL kernel, i.e.  it does not contain the gluon reggeization).
Later on it will be convenient to rewrite the reggeization of the
gluon (third term on the right-hand-side): putting $\ln (s/\bm r^2) =
\ln (s/s_0) + \ln (s_0/\bm r^2)$ we combine the $\ln (s/s_0)$ piece
with the real part of the BFKL kernel: in the sum of both terms the
infrared singularity in the limit $\bm r - \bm r' \to 0$ drops out.
Note the factorization of eq.~(\ref{eq:svil}): as indicated after
(\ref{eq:regge}), this feature is a general consequence of Regge
theory and (with suitable generalizations of the $t$-channel gluon
propagators) is expected to hold in arbitrary order perturbation
theory. For the impact factor we state the general definition:
\begin{equation}
  \label{eq:ifdef1}
  \Phi_A = \frac{\delta^{ab}}{\sqrt{N_c^2 -1}}  \sum_{A'}
\int \left\langle\Gamma_{A \to A'}^a 
  \Gamma_{A \to A'}^{b\;*}\right\rangle 
  d\phi_{A'} \frac{sd\beta_r}{2\pi}
\end{equation}
The average symbol $\langle\rangle$ stands for the sum over color and
helicity in the intermediate state $A'$. In the following we will use
the short hand notation
\begin{equation}
  \label{eq:ifdef}
  \Phi_A = \sum_{A'} \int |\Gamma_{A \to A'}|^2 
  d\phi_{A'} \frac{sd\beta_r}{2\pi},
\end{equation}
In particular, in the square of the matrix elements we have included
the colour projector $\delta^{ab}/\sqrt{N_c^2-1}$.

In this paper we are interested in the NLO corrections to the photon
impact factor. The strategy follows from the discussion outlined in
this section.  The NLO corrections can be divided into the $q\bar{q}$
and $q\bar{q}g$ intermediate states (named virtual and real
corrections, respectively).  In the former case we need the NLO
corrections to the $\gamma^* \to q\bar{q}$ vertex, $\Gamma_{\gamma^*
  \to q\bar{q}}^{(1)}$.  They contain infrared divergences (in
dimensional regularization: poles in $\epsilon$) and are listed in
\cite{BGQ}.  An additional (infrared divergent) contribution comes
from the gluon trajectory function in the third term on the
right-hand-side of (\ref{eq:svil}).  For the $q\bar{q}g$ contribution
we need the vertex $\Gamma_{\gamma^*\to q\bar{q}g}$, in the Born
approximation: since, by definition, the gluon lies in the
fragmentation region of the virtual photon, we have to divide the
phase space of the produced gluon.  In section~\ref{sec:NLOIF} we
present a detailed discussion of this rather subtle issue: we will
show that the separation of the fragmentation region introduces the
scale $s_0$, in the last term of (\ref{eq:svil}). The fragmentation
region contains soft and collinear divergences. When combining them
with the infrared divergences of the $q\bar{q}$ state (including
those from the gluon trajectory function), the infrared divergences
cancel \cite{FM}. It is the main purpose of this paper to find a
separation of the fragmentation from the central region which allows
to define infrared finite combinations of virtual and real corrections
to the impact factor.

\section{Singular Terms of the Virtual Corrections}
\label{sec:singular-virtual}

\begin{figure}[htbp]
  \begin{center}
    \epsfig{file=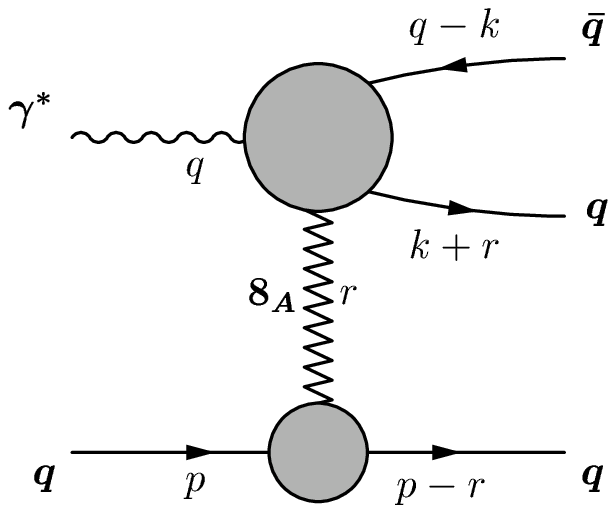}
    \caption{Kinematics of the process $\gamma^*+q' \to q\bar q + q'$.
      \label{fig:qqkin}}
  \end{center}
\end{figure}

\noindent
For future reference, and in order to define our notation we briefly
summarize the Born level impact factor and list the divergent terms of
the virtual corrections.  The amplitude $T_{\gs q}$ for the scattering
$\gamma^*+q' \to q\bar q + q'$ (Fig.~\ref{fig:qqkin}) has been
calculated in \cite{BGQ} to next-to-leading order in $\as$.  We write
$T_{\gs q}$ as an expansion in the strong coupling $g$
\begin{equation}
  \label{eq:NLOgsq}
  T_{\gs q} = g^2 T_{\gs q}^{(0)} + g^4 T_{\gs q}^{(1)}\, .
\end{equation}

The momenta are labelled as in Fig.~\ref{fig:qqkin}.  The kinematical
variables we use are the centre-of-mass energy $s$ of the incoming
virtual photon and the incoming quark, the virtuality of the photon,
$Q^2$, the Bjorken variable $x=Q^2/2p\cdot q$, the momentum transfers
$t=r^2$, $t_a=k^2$, $t_b=(q-k-r)^2$, and the invariant mass $M^2$ of
the outgoing $q\bar q$-system.  In addition we use a Sudakov
decomposition of the momentum $k$ with respect to the light cone
momenta $q' = q-xp$ and $p$ with $2p\cdot q' = s$ , $k = \alpha q' +
\beta p + k_\perp$.  Two dimensional transverse momenta are denoted as
$\bm k^2 = -k_\perp^2$.

The Born impact factor is obtained from $T^{(0)}_{\gs q}$. To leading
order we may write
\begin{equation}
  \label{eq:bornqq}
  T_{\gs q}^{(0)} = \Gamma_{\gamma^* \to q\bar{q}}^{(0),a}\; \frac{2s}{t} \;
  \Gamma_{qq}^{(0),a} 
\end{equation}
with the Born level vertices
\begin{align}
  \label{eq:qvertexborn}
  \Gamma_{qq}^{(0),a} &= \frac{1}{s} \bar{u}(p-r, \,\lambda_{q'}) \not\!{q}'
  \lambda^a u(p,\,\lambda_q)\,, \\
  \label{eq:gsvertexborn}
  \Gamma_{\gamma^* \to q\bar{q}}^{(0),a} &= -iee_f \left(\frac{H_T^a}{st_a} -
    \frac{\bar{H}_T^a}{st_b} \right)\, ,
\end{align}
where
\begin{eqnarray}
  H_{T}^a &=& \bar{u}(k+r, \lambda) \not\!{p} \not\!{k} \not\!{\varepsilon}\; 
  \lambda^av(q-k,\lambda') \; , \\
  \bar{H}_T^a&=&
  \bar{u}(k+r, \lambda) \not\!{\varepsilon}\;  
  (\not\!{q}-\not\!{k}-\not\!{r})\; 
  \not\!{p}\; \lambda^av(q-k,\lambda') \;
\end{eqnarray}
and $\lambda^a$ are the generators of the colour group $SU(N_c)$.
Squaring the Born level vertices (\ref{eq:qvertexborn}) and averaging
(summing) over incoming (outgoing) colours and helicities we obtain
for the coupling to the reggeized gluons with colours $(a,b)$ ,
contracted with the colour singlet projector (cf.
eqs.~(\ref{eq:ifdef1}) and (\ref{eq:ifdef})):
\begin{equation}
  \label{eq:sq_qvertex}
  |\Gamma^{(0)}_{qq}|^2 = \frac{\sqrt{N_c^2-1}}{2N_c}\,. 
\end{equation}
The $\gs\to q\bar q$-vertex $|\Gamma^{(0)}_{\gs\to q\bar q}|^2 =
|\Gamma^{(0)}_{\gs\to q\bar q}(\bm k, \alpha)|^2$ for a longitudinally
polarized $\gs$ coupling to a quark flavour $f$ reads
\begin{equation}
  \label{eq:sq_gsvertex_L}
  |\Gamma^{(0)}_{\gs\to q\bar q}|^2_{L} =
  4\sqrt{N_c^2-1}e^2 e_f^2 \alpha^3 (1-\alpha)^3 Q^2
  \left(\frac{1}{D(\bm k)} - \frac{1}{D(\bm k+\bm r)}\right)^2 \,,
\end{equation}
where $D(\bm k) = \bm k^2 + \alpha(1-\alpha)Q^2$.  Here the exchanged
reggeized gluons must be in the colour singlet.  For the sum over the
two transverse photon polarizations we have
\begin{equation}
  \label{eq:sq_gsvertex_T}
  |\Gamma^{(0)}_{\gs\to q\bar q}|^2_{T} =
  2\sqrt{N_c^2-1} e^2 e_f^2 \alpha (1-\alpha) 
  \left[\alpha^2 + (1-\alpha)^2 \right]
  \left(\frac{\bm k}{D(\bm k)} 
    - \frac{\bm k+\bm r}{D(\bm k+\bm r)}\right)^2 \,.
\end{equation}
The LO impact factor follows from the definition in eq.\ 
(\ref{eq:ifdef}), from the two-particle phase space measure
\begin{equation}
   \label{eq:qqps}
  d\phi_{q\bar q} \frac{sd\beta_r}{2\pi} = 
  \frac{d\alpha}{2 \alpha (1-\alpha)} 
  \frac{d^{D-2}\bm k}{(2\pi)^{D-1}}     
\end{equation}
and from the Born level vertices~(\ref{eq:sq_gsvertex_L},
\ref{eq:sq_gsvertex_T}).  Expanding $\Phi_{\gs}$ in the strong
coupling, 
\begin{equation}
  \label{eq:ifexpansion}
  \Phi_{\gs} = g^2 \Phi_{\gs}^{(0)} + g^4 \Phi_{\gs}^{(1)},
\end{equation}
we write the Born level impact factor as 
\begin{equation}
  \label{eq:bornif}
  \Phi_{\gs;T,L}^{(0)} =
  \int\frac{d^{D-2}\bm k}{(2\pi)^{D-1}}   
  \frac{d\alpha}{2 \alpha (1-\alpha)} |\Gamma_{\gs\to q\bar q}^{(0)}|_{T,L}^2
  \equiv
  \int d^{D-2}\bm k \, d\alpha \,\I_{2;T,L}(\alpha,\bm k;\bm r,Q)\, .
\end{equation}
The last equation defines the integrand of the LO impact factor:
\begin{equation}
\label{eq:bornifl}
\I_{2;L} (\alpha,\bm k;\bm r,Q) =\frac{2e^2 e_f^2\sqrt{N_c^2-1}}{(2\pi)^{D-1}} 
 \alpha^2 (1-\alpha)^2 Q^2
 \left(\frac{1}{D(\bm k)} - \frac{1}{D(\bm k+\bm r)}\right)^2
\end{equation}
\begin{equation}
\label{eq:bornift}
\I_{2;T} (\alpha,\bm k;\bm r,Q) =\frac{e^2 e_f^2 \sqrt{N_c^2-1}}{(2\pi)^{D-1}}
\left[\alpha^2 + (1-\alpha)^2 \right]
\left(\frac{\bm k}{D(\bm k)} 
    - \frac{\bm k+\bm r}{D(\bm k+\bm r)}\right)^2
\end{equation}
Because of the symmetry of the squared matrix element under the exchange 
$q\leftrightarrow\bar q$, namely $\alpha\leftrightarrow1-\alpha$, 
$\bm k\leftrightarrow-\bm r-\bm k$, we have the relation
$\I_2 (\alpha, \bm k;\bm r,Q)=\I_2 (1-\alpha,-\bm r-\bm k;\bm r,Q)$.

Turning to the NLO part of (\ref{eq:NLOgsq}), we split $T_{\gs q}$ into a finite part $F$
and a divergent part $D$ as
\begin{equation}\label{eq:T1div}
  T_{\gs q}^{(1)} = F_{\gamma^*q} + \frac{c_\Gamma}{(4\pi)^{2-\epsilon}} 
T_{\gs q}^{(0)} D\,,\qquad
  c_\Gamma\equiv\frac{\Gamma(1+\epsilon)\Gamma^2(1-\epsilon)}{\Gamma(1-2\epsilon)}\,. 
\end{equation}
The divergences are given in dimensional regularization and appear as
poles when the space-time dimension $D = 4 - 2\epsilon$ approaches
$D=4$.  Here we explicitly include the divergent contribution from
self-energy insertions on all four external quark lines
\begin{equation}
\label{eq:SE}  
D_{\mathrm{SE}} = -2 C_F \left(\frac{1}{\epsilon_{\mathrm{UV}}} 
    - \frac{1}{\epsilon}\right)\, ,
\end{equation}
labelling poles from ultraviolet divergences explicitly by
$\epsilon_{\mathrm{UV}}$. As usual, the $SU(N_c)$ invariants are
denoted by $C_A=N_c$ and $C_F = (N_c^2 - 1)/(2N_c)$.

The divergent part can then be written as the sum of three terms
\begin{equation}
  \label{eq:virtdiv}
  D = D_{\gamma^*\to q\bar q} + D_{\omega} + D_{qq}\,,
\end{equation}
following the expansion of the Regge ansatz \eqref{eq:regge} in powers
of $\alpha_s$. The three terms denote the divergent parts of the
vertex $\Gamma_{\gamma^*\to q\bar q}^{(1)}$, of the Regge trajectory
of the gluon, $\omega^{(1)} (t)$, which is accompanied by the large
leading logarithms in $s$, and of the vertex $\Gamma_{qq}^{(1)}$,
respectively. They have the form:
\begin{equation}
  \label{eq:gsdiv}
  D_{\gamma^*\to q\bar q} = 
  - \frac{2C_F}{\epsilon^2} 
  - \frac{3C_F}{\epsilon} 
  + \frac{\beta_0}{\epsilon_{\mathrm{UV}}} 
  + \frac{2C_F}{\epsilon} 
  \ln\left(-M^2\right) +
   \frac{C_A}{\epsilon} 
    \left(
      \ln\frac{\alpha (1-\alpha)\, t}{M^2} 
      + \ln\frac{\hfill s_0}{-t}
    \right)\,,
\end{equation}
\begin{equation}
  \label{eq:omegadiv}
  D_{\omega} = 
  \frac{C_A}{\epsilon}\left[
    \ln\frac{\hfill s}{s_0} + \ln\frac{-s}{\hfill s_0}\right]
\end{equation}
and
\begin{equation}
  \label{eq:qqdiv}
  D_{qq} = 
  - \frac{2C_F}{\epsilon^2} 
  - \frac{3C_F}{\epsilon} 
  + \frac{\beta_0}{\epsilon_{\mathrm{UV}}} 
  + \frac{2C_F}{\epsilon}\ln\left(-t\right) 
  + \frac{C_A}{\epsilon}\ln\frac{\hfill s_0}{-t}\;.
\end{equation}
Here, we have separated infrared and ultraviolet divergences, making
use of the formulae given in \cite{BGQ}. Note that, compared to the
Regge ansatz (\ref{eq:regge}), we have changed the energy scale. In
(\ref{eq:omegadiv}), we have introduced the energy scale $s_0$:
instead of $\ln s/(-t)$ we write $\ln s/(-t)=\ln s/s_0 + \ln s_0/(-t)$
and absorb the second term into $D_{\gamma^* \to q\bar{q}}$ and
$D_{qq}$ in (\ref{eq:gsdiv}, \ref{eq:qqdiv}). Let us first consider
the infrared singular terms.  For the virtual NLO corrections to the
photon impact factor we start from
\begin{equation}
  \label{eq:Phi1virt}
 \Phi_{\gs}^{(1,{\mathrm virtual})} = \frac{\delta_{ab}}{\sqrt{N_c^2 -1}}
\int \frac{d^{D-2}\bm k}{(2\pi)^{D-1}}   
  \frac{d\alpha}{2 \alpha (1-\alpha)}
\Gamma_{\gs\to q\bar q}^{(1),a}\Gamma_{\gs\to q\bar q}^{(0),b\;*}
 d\phi_{q\bar q} \frac{sd\beta_r}{2\pi} +\text{c.c.} \,.
\end{equation}
Making use of eqs.~(\ref{eq:T1div}) and (\ref{eq:gsdiv}) --- the latter 
one without the UV pole which we will discuss later --- we can immediately
deduce the IR divergent part of the virtual correction to the impact factor:
\begin{align}\nonumber
 &\left.\Phi_{\gs}^{(1,{\mathrm virtual})}\right|^{\mathrm IR divergent} =
 \frac{c_\Gamma}{(4\pi)^{2-\epsilon}}\Phi_{\gs}^{(0)}\,
 \left[D_{\gamma^*\to q\bar q}^{\mathrm IR} + \text{c.c.}\right] \\
&=\int d\bm k\int_0^1d\alpha\,\I_2(\alpha, \bm k; \bm r,Q)
 \frac{c_\Gamma}{(4\pi)^{2-\epsilon}}\nonumber \\
&\quad\times\biggl\{
 \frac{C_A}{\epsilon}\left[2\log s_0+2\log\alpha(1-\alpha)-2\log M^2\right]
 +\frac{C_F}{\epsilon}\left[-\frac4{\epsilon}+4\log M^2-6\right]\biggr\}\,.
 \label{eq:Phi1virtSing}
\end{align}
Below we will show that these divergent terms cancel against the real
corrections.

Turning to the ultraviolet divergent pieces in $D_{\gamma^*\to q\bar
  q}$ and $D_{qq}$, they are proportional to $\beta_0 = (11 N_c - 2
n_f)/6$, and in \cite{BGQ} it has been shown that they are removed by
the renormalization of the strong coupling. It is instructive to
repeat the argument in more detail. The ultraviolet part of
(\ref{eq:gsdiv}), including constants, is obtained by adding, in
\cite{BGQ}, eqs.~(56), (57), (61), (62), (67), plus one half of
eq.~(60):
\begin{equation}
\label{eq:uvdivergent}
\frac{g^4 c_{\Gamma}}{(4\pi)^{2-\epsilon}} 
\Gamma_{\gs\to q\bar q}^{(0), a}
\left[ \frac{(\bm r^2)^{-\epsilon} }{\epsilon_{UV}}
\left(\frac{14N_c}{6} - \frac{1}{3}n_f -
\frac{1}{2N_c}\right) -\frac{C_F}{\epsilon_{UV}} \right]
\end{equation}
Except for the last term which is due to the wave function renormalization 
of the massless outgoing quarks, all ultraviolet poles are proportional to 
$(\bm r^2)^{-\epsilon}$. It will be convenient to slightly modify the last 
term: instead of using, for the renormalization of the outgoing massless 
quarks, the decomposition (\ref{eq:SE}), we write 
\begin{equation}
\label{eq:SEP}
D'_{\mathrm{SE}} = -2 C_F 
\left(\frac{(\bm r^2)^{-\epsilon}}{\epsilon_{\mathrm{UV}}} 
  - \frac{(\bm r^2)^{-\epsilon}}{\epsilon}\right).
\end{equation}    
This allows to simplify (\ref{eq:uvdivergent}):
\begin{equation}
\frac{g^4 c_{\Gamma}}{(4\pi)^{2-\epsilon}}
\Gamma_{\gs\to q\bar q}^{(0), a}
\frac{(\bm r^2)^{-\epsilon}}{ \epsilon_{UV}}
\beta_0 
\end{equation}
When addressing, in the next step of our program, the calculation of
the finite terms of the virtual corrections, we will have to remember
that the change (\ref{eq:SEP}) in the wave function renormalization
introduces a new finite term
\begin{equation}
  \label{eq:newfiniteSEterm}
  -\frac{g^4}{(4\pi)^2} 
  \Gamma_{\gs\to q\bar q}^{(0), a}\,
  C_F \ln(\bm r^2).
\end{equation}
In order to remove the UV divergences we have to renormalize, i.e. we
add the Born approximation with the replacement:
\begin{equation}
g \to \frac{Z_1g_{\mu} \mu^{\epsilon}}{Z_2 \sqrt{Z_3}} = 
g_{\mu} \mu^{\epsilon} \left[1 - \frac{\alpha_s}{4\pi}\beta_0 
\left(\frac{1}{\epsilon_{UV}} - \gamma_E + \ln (4\pi) \right) + ...\right] 
\end{equation}
with the renormalization constants taken in the $\overline{\rm{MS}}$
scheme. When both contributions are added, the ultraviolet poles
cancel, and we are left with
\begin{equation}
\label{eq:vertexrenorm}
g^2 \Gamma_{\gamma^* \to q\bar{q}}^{(1;\rm{UV})}= 
- \frac{\alpha_s}{4\pi} \Gamma_{\gamma^* \to q\bar{q}}^{(0)} \,
\beta_0 \ln \frac{\bm r^2}{\mu^2}.
\end{equation}   
Note that all dependence upon the renormalization scale $\mu$ is
contained in the logarithm which introduces the scale dependence of
the strong coupling constant, $\alpha_s(\bm r^2)$.  Correspondingly,
in the photon impact factor the $\mu$-dependence is contained in the
contribution:
\begin{equation}
\label{eq:impactrenorm}
g^2 \Phi_{\gamma^*}^{(1;\rm{UV})} = - \frac{\alpha_s}{2 \pi}
\Phi_{\gamma^*}^{(0)}
\beta_0 \ln \frac{\bm r^2}{\mu^2}
\end{equation}
as it is required by the renormalization group equation. Our derivation 
might suggest that $\bm r^2$ is the `natural' momentum scale of $\alpha_s$
in the photon impact factor: however, it is clear that we can write 
\begin{equation}
\label{eq:impactrenormp}
g^2 \Phi_{\gamma^*}^{(1;\rm{UV})} = - \frac{\alpha_s}{2 \pi}
\Phi_{\gamma^*}^{(0)}\left(
  \beta_0 \ln \frac{Q^2}{\mu^2}
  + \beta_0 \ln \frac{\bm r^2}{Q^2}
\right)
\end{equation} 
i.e. by simply redefining the $\mu$-independent part of the impact factor 
we can switch from one momentum scale of $\alpha_s$, $\alpha_s(\bm r^2)$, 
to another scale, $\alpha_s(Q^2)$.        

Stating that (\ref{eq:impactrenorm}) is the only $\mu$-dependent term
in the impact factor implies that the $\mu$-dependence drops out of
all the remaining contributions to the impact factor.  In fact, we
silently dropped a $\mu$ already in eq.~(\ref{eq:newfiniteSEterm}),
anticipating the scale invariance of the finite terms which are left after 
the cancellation of the infrared divergencies of real and virtual 
corrections. This can be understood as follows. 
Without having it written explicitly in \cite{BGQ}, the infrared
divergences in dimensional regularization appear as poles in
$\epsilon$, accompanied by the power of some dimensionful scale, $s_i$.
In a one-loop calculation the sum of infrared divergent terms from
real (R) and virtual (V) corrections generally reads
\begin{equation}
  \label{eq:irtermsgeneral}
  T_{\rm{IR}} = \sum_i A_i \frac{(s_i)^{-\epsilon}}{\epsilon^2} + 
B_i \frac{(s_i)^{-\epsilon}}{\epsilon}. 
\end{equation}
Adding real and virtual corrections, the coefficients $A_i$ and $B_i$
are sums as well, 
\begin{equation}
  A_i = A_i^V + A_i^R, \qquad B_i = B_i^V + B_i^R.
\end{equation}
Expanding (\ref{eq:irtermsgeneral}) in $\epsilon$ we obtain
\begin{equation}
  \label{eq:irtermsexp}
  T_{\rm{IR}} = \sum_i \frac{A_i}{\epsilon^2}  - \frac{A_i}{\epsilon} \ln s_i
  + \frac{A_i}{2} \ln^2 s_i
  + \frac{B_i}{\epsilon}  - B_i\ln s_i. 
\end{equation}
Introducing the scale $\mu$ with the renormalized coupling as a factor
$\mu^{2\epsilon}$ and reexpanding results in 
\begin{align}
  \label{eq:irmuexp}
  T_{\rm{IR}} =& \sum_i
  \left(  \frac{1}{\epsilon^2} + \frac{\ln\mu^2}{\epsilon}  
    + \frac{\ln^2 \mu^2}{2} \right) A_i
  - \left( \frac{1}{\epsilon} + \ln \mu^2 \right) A_i \ln s_i
  + \frac{A_i}{2} \ln^2 s_i \nonumber \\
  &+ \left( \frac{1}{\epsilon} + \ln \mu^2 \right) B_i  
  - B_i \ln s_i. 
\end{align}
Further below (section~\ref{sec:finite}) we will show that in the sum
of virtual and real corrections all $\epsilon$-poles cancel: this
means that $A_i = 0$ and $\sum_i B_i = 0$.  Hence, we see that the
cancellation of infrared divergences implies that also the $\ln \mu^2$
terms cancel.  This allows us to write e.g.\ the logarithm in
(\ref{eq:newfiniteSEterm}) with a dimensionful argument.  We did the
same with the results of diagram 14 in \cite{BGQ} after expanding
divergent terms of the form $s_i^{-\epsilon}$.

\section{Real Corrections}
\label{sec:real}

\begin{figure}[htbp]
  \begin{center}
    \epsfig{file=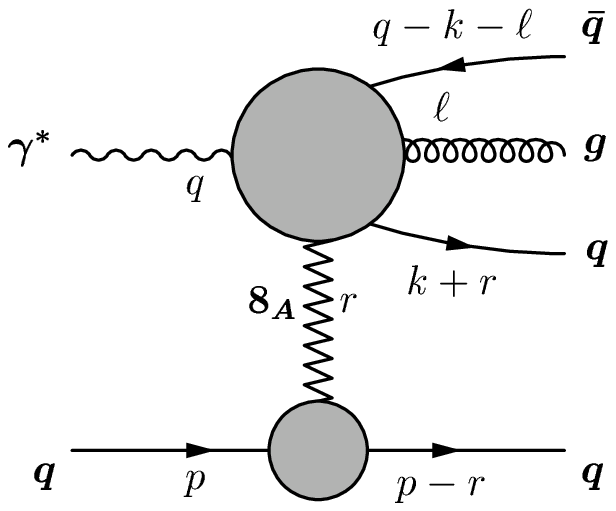}
    \caption{Kinematics of the process $\gamma^*+q' \to q\bar qg + q'$.
      \label{fig:qqgkin}}
  \end{center}
\end{figure}

\begin{figure}[h]
  \begin{center}
    \epsfig{file=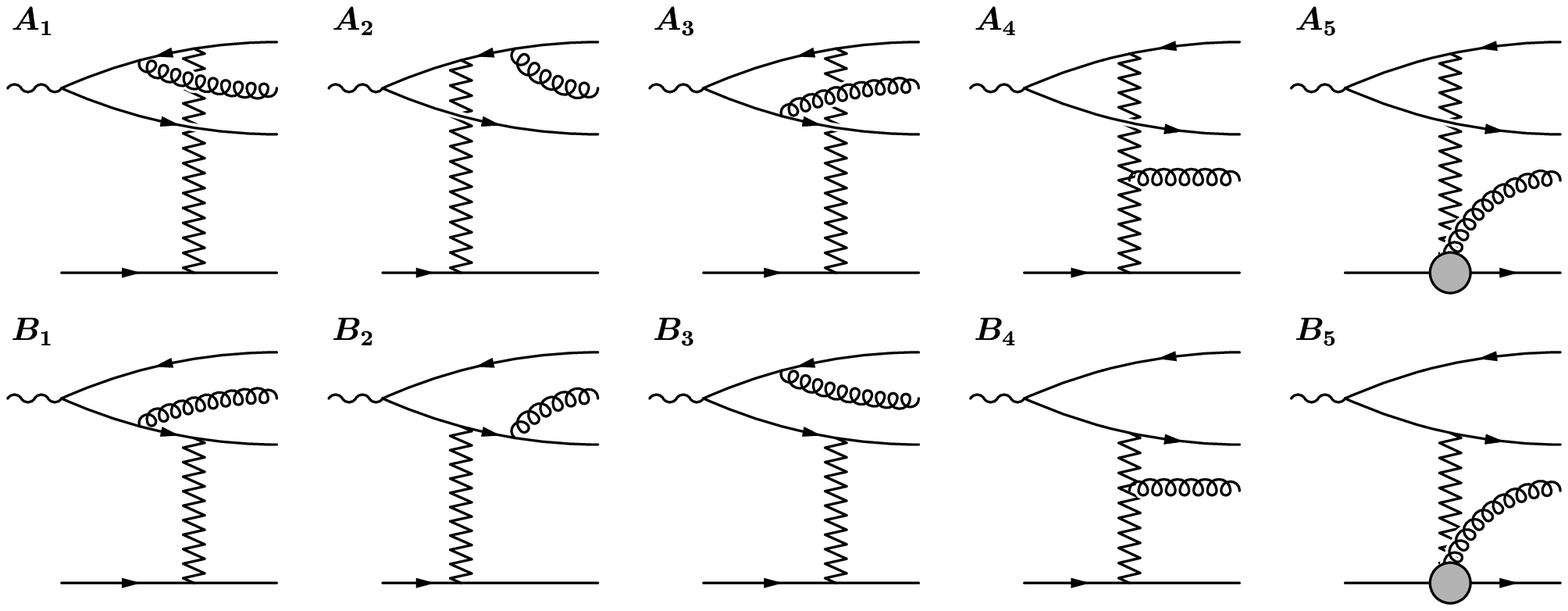,width=\textwidth}
    \caption{Feynman diagrams for $\gamma^*+q' \to q\bar qg + q'$.
      \label{fig:diags}}
  \end{center}
\end{figure}

\noindent
We now turn to the real corrections. We study the process $\gs +q \to
(q{\bar q}g) + q$ with transverse photon polarization
(Fig.\ref{fig:qqgkin}); the case of the longitudinal photon has been
studied in \cite{BGK}, and we follow the procedure outlined in this
earlier paper.  The diagrams are shown in Fig.~\ref{fig:diags}. We use
the Sudakov decomposition of the gluon momentum $\ell = \alpha_\ell q'
+ \beta_\ell p + \ell_\perp$ and introduce the following abbreviations
:
\begin{align}
  \label{eq:bardefvars}
  \alpha_1 &\equiv \alpha\,,\\
  \bar \alpha_1 &\equiv (1-\alpha-\alpha_\ell)\,,\\
  \alpha_2 &\equiv (1-\alpha)\,,\\
  \bar \alpha_2 &\equiv (\alpha+\alpha_\ell).
\end{align}

The propagator denominators, that occur in the calculation, have the
following form:
\begin{align}
D_1&=(k+\ell+r-q)^2 \nonumber\\
&={}-\bar{\alpha}_1 Q^2 - (\bm k + \bm\ell + \bm r)^2 -
\frac{\bar{\alpha}_1}{\alpha_1}\, (\bm k + \bm r)^2 -
\frac{\bar{\alpha}_1}{\alpha_\ell} \, \bm\ell^2\label{eq:D1}\\
D_2&=(k+r-q)^2 \nonumber\\
&={}-\alpha_2 \, Q^2 - \frac{1}{\alpha_1} \,
(\bm k + \bm r)^2\label{eq:D2}\\
D_3 &=(k+\ell+r)^2\nonumber\\
&={}-(\bm k + \bm\ell + \bm r)^2+ 
\frac{\bar{\alpha}_2}{\alpha_1} \,
(\bm k + \bm r)^2 + \frac{\bar{\alpha}_2}{\alpha_\ell}\, \bm\ell^2 
\nonumber\\
&={}\frac{1}{\alpha_\ell \, \alpha_1}\, (-\alpha_\ell \,
(\bm k + \bm r) + \alpha_1\, \bm\ell )^2\\
D_4&=(k-q)^2\nonumber\\
&={}-\bm{k}^2 + \frac{\alpha_2}{\bar{\alpha}_1} \, 
(\bm k + \bm\ell)^2
+ \frac{\alpha_2}{\alpha_\ell} \, \bm\ell^2 \nonumber \\
&={}\frac{1}{\alpha_\ell \, \bar{\alpha}_1} \,
(\alpha_\ell \, (\bm k + \bm\ell) + \bar{\alpha}_1 \, \bm\ell)^2\\
D_5&=(\ell-r)^2\nonumber\\
&={}-\alpha_\ell \, Q^2 - \frac{\alpha_\ell}{\alpha_1} \,
(\bm k + \bm r)^2 - \frac{\alpha_\ell}{\bar{\alpha}_1} \, 
(\bm k + \bm\ell)^2 -
(\bm\ell-\bm r)^2\\
D_6&=k^2\nonumber\\
&={}-\alpha_1 \, Q^2 - \bm{k}^2 - \frac{\alpha_1}{\bar{\alpha}_1}
\, (\bm k + \bm\ell)^2  - \frac{\alpha_1}{\alpha_\ell} \, \bm\ell^2\\
D_7&=(k+\ell)^2\nonumber\\*
&={}-\bar{\alpha}_2 \, Q^2 - \frac{1}{\bar{\alpha}_1} \,
(\bm k + \bm\ell)^2\label{eq:D7}\,.
\end{align} 

\noindent
We present the results for $|{\mathcal M}_{q{\bar q}g}|^2$, the squared matrix element, averaged (summed)
over the helicities and  colors of the incoming (outgoing) quarks and
summed over equal photon polarizations; thereby we made use of ~{$\sum_{i=j=1,2}\,
(\ep_T^i\, k_\perp) (\ep^j_T\, \ell_\perp) = {\bm k} {\bm \ell}$}.
Our results can be simplified by expressing them in terms of 
the matrix elements for longitudinal photons; for this reason we also recall 
the results for longitudinal photon polarization \cite{BGK}.
The matrix elements for the process $\gamma^*q \to (q\bar{q}g)q$ can be 
written in the following factorized form: 
\begin{equation}
|{\mathcal M}_{q{\bar q}g}|^2 = \frac{1}{N
_c^2 - 1}\; g^2 \, \delta^{ab} \; |\Gamma^{(0)}_{\gamma^* \to q{\bar
  q}g}|^2 \; \left( \frac{2\,s}{t} \right)^2\; \delta^{ab} \; |\Gamma^{(0)}_{qq}|^2,
\end{equation}
and by using eq.(\ref{eq:sq_qvertex}) we extract the squared vertex 
$|\Gamma^{(0)}_{\gamma^* \to q{\bar
    q}g}|^2$ (summed over helicities and color in the $q\bar{q}g$ state).
Following the notation used in \cite{BGK} we list our results 
in correspondence with the Feynman diagrams:
\begin{equation}
\label{eq:m2l}
|\Gamma^{(0)}_{\gamma^* \to q{\bar
    q}g}|^2_L = 4 \,e^2e_f^2g^4\,\sqrt{N_c^2 - 1}\;
 Q^2 \, \sum_{i,j=1}^{5} \; \left( \A\A_{ij}^L \,+\,\A\B_{ij}^L 
    \,+\, \B\A_{ij}^L \,+\,  \B\B_{ij}^L\right), 
\end{equation}
\begin{equation}
\label{eq:m2t}
|\Gamma^{(0)}_{\gamma^* \to q{\bar
    q}g}|^2_T = e^2e_f^2g^4\,\sqrt{N_c^2 - 1}\;
\sum_{i,j=1}^{5} \; \left( \A\A_{ij}^T \,+\,\A\B_{ij}^T 
    \,+\, \B\A_{ij}^T \,+\,  \B\B_{ij}^T\right); 
\end{equation}
\begin{align}
\A\A_{11}^T =&{}\;  C_F \; \frac{4 }{{D_1}^2\,{D_2}^2\,
      } \alpha_1\,{\bar \alpha}_1\,
     ( 1 - \ep ) \,
     \bigg( \klr^2\,\kr^2 \\& \nonumber + 
       \frac{1}{\alpha_1^2} {\bar \alpha}_1\,\Big( \kr^2 + 
            \alpha_1\,Q^2 \Big) \,
          \Big( 2\,\alpha_1\,\krklr 
             + {\bar \alpha}_1\,( \kr^2 + 
               \alpha_1\,Q^2 )  \Big)
            \bigg) \, \\& \nonumber + 
  \frac{2\,\kr^2}
   {{\alpha_1}^2\,} \,\A\A_{11}^L\\[.3cm] 
\A\A_{12}^T =&{}\;- \frac{1}{N_c}\;  \frac{2 }{D_1\,{D_2}^2\,
     D_4\, } \bigg[ {\bar \alpha}_1\,\kr^2\,
        ( \bm{k r} + \alpha_2 \bm{r}^2
           )  + \alpha_1\,\alpha_2\,{\bar \alpha}_1\,
        \rkr \,Q^2\ \\& \nonumber
        + ( 1 - \ep ) \,
        \bigg( \Big(  - \alpha_1\,{\bar \alpha}_1\,
                \kklr   - 
             {{\bar \alpha}_1}^2\,\kkr + 
             \alpha_1\,\alpha_2\,\klklr \\& \nonumber \quad +
             \alpha_2\,{\bar \alpha}_1\,\klkr \Big) \,
           \kr^2 \\& \nonumber - \alpha_1\,{\bar \alpha}_1\,
           ( {\bar \alpha}_1\,\kkr - 
             \alpha_2\,\klkr ) \,Q^2\
          \bigg)  \bigg] \, + 
  \frac{2\,\kr^2}
   {{\alpha_1}^2\,} \,\A\A_{12}^L\\[.3cm]
\A\A_{13}^T =&{}\; C_F\; \frac{4 }{
    {\alpha_1}^2\,\alpha_{\ell}\,{D_1}^2\,D_2\,
    D_3\, } \,{\bar \alpha}_1 \,
    \bigg[- {\alpha_1}^3\,( {\bar \alpha}_2\,\alpha_{\ell}\,
          \klr^2 - {{\bar \alpha}_1}^2\, \bm{\ell}^2 ) \,
       Q^2 \\& \nonumber + {\alpha_1}^2\,\krklr\,
       \Big( \alpha_{\ell}\,\klr^2 + 
         {\bar \alpha}_1\, \bm{\ell}^2 - 
         {\bar \alpha}_1\,\alpha_{\ell}\,Q^2 \Big)  + 
      \alpha_{\ell}\,\kr^2\, \\& \nonumber 
       \Big( - \alpha_1\,{\bar \alpha}_2\,\klr^2\
             + {{\bar \alpha}_1}^2\,\kr^2 + 
         \alpha_1\,{\bar \alpha}_1\,\krklr - \\& \nonumber
         \alpha_1\,{{\bar \alpha}_1}^2\,
          ( - \frac{ \bm{\ell}^2}{\alpha_{\ell}}   - 
            \alpha_1\,Q^2 )  \Big) \\& \nonumber+ 
      \alpha_1\,( 1 - \ep ) \,
       \bigg( - {\alpha_1}^2\,{{\bar \alpha}_1}^2\, \bm{\ell}^2\,
          Q^2  - \alpha_1\,\alpha_{\ell}\,\krklr\,
          \Big( \alpha_2\,\klr^2  \\& \nonumber- 
            4\,{\bar \alpha}_1\,\krklr +
            \frac{\alpha_1\,{\bar \alpha}_1\, \bm{\ell}^2}
             {\alpha_{\ell}} - {{\bar \alpha}_1}^2\,Q^2 \Big) \\& \nonumber
          + \alpha_{\ell}\,\kr^2\,
          \Big( ( -2\,\alpha_1\,{\bar \alpha}_1  + 
               \alpha_2\,{\bar \alpha}_2 ) \,\klr^2 - 
            {{\bar \alpha}_1}^2\,\kr^2 \\& \nonumber \quad + 
            {\bar \alpha}_1\,( -\alpha_1 + {\bar \alpha}_1 - 
               {\bar \alpha}_2 ) \,\krklr + 
            \alpha_1\,{{\bar \alpha}_1}^2\,
             ( - \frac{ \bm{\ell}^2}{\alpha_{\ell}}   - 
               Q^2 )  \Big)  \bigg)  \bigg] \,\\[.3cm] 
\A\A_{14}^T =&{}\; - N_c\;  \frac{1}{D_1\,
     {D_2}^2\,D_5\, }\,\bigg[ {{\bar \alpha}_1}^2\,\kr^2\,
        ( -\kr^2 + 
          \frac{\alpha_1\, \bm{\ell}^2}{\alpha_{\ell}} - 
          \alpha_1\,Q^2 )   \\& \nonumber +
       \frac{2\,\alpha_{\ell} }{\alpha_1}\,( 1 - \ep ) \,
          \bigg( \alpha_1\,{\bar \alpha}_1\,
             \Big( \kr^2 + \alpha_1\,Q^2\
               \Big) \,( \,2 \klkr + \rkr  + 
               {\bar \alpha}_1\,Q^2 ) \\& \nonumber\quad + 
            \kr^2\,\Big( {\alpha_1}^2\,\klklr + 
               {{\bar \alpha}_1}^2\,\kr^2 + 
               \alpha_1\,{{\bar \alpha}_1}^2\,Q^2 \Big) \
            \bigg) \\& \nonumber+ 
       \alpha_1\,{\bar \alpha}_1\,\kr^2\,
        \Big(- \kkl + \lmr \,\klr - \bm{r}^2 
           \Big) \\& \nonumber + 
       {\bar \alpha}_1\,( -\kr^2 - 
          \alpha_1\,Q^2 ) \,
        \Big( \alpha_{\ell}\,(\,2 \klkr + 
          \rkr) \\& \nonumber \quad+ 
          {\bar \alpha}_1\,( 2\,\bm{\ell} - \bm{r})\kr
               \Big)  \bigg] \, \\& \nonumber + 
  \frac{2\,\kr^2}{{\alpha_1}^2\,} \,\A\A_{14}^L\\[.3cm]
\A\A_{15}^T =&{}\;N_c\;  \frac{2\,}{
    \alpha_1\,\alpha_{\ell}\,D_1\,{D_2}^2\,
    D_5}\,( {\alpha_1}^2 + {\alpha_2}^2 )\,
    {{\bar \alpha}_1}^2\,\kr^2 \, \bm{r}^2\\
\A\A_{22}^T =&{}\;  \frac{ 2\, ( {\alpha_1}^2 + {\alpha_2}^2\
        ) \,\kr^2  }{
    {\alpha_1}^2\,{\alpha_2}^2\,}\,\A\A_{22}^L\\[.3cm]
\A\A_{23}^T =&{}\;- \frac{1}{N_c} \; \frac{2\,}{D_1\,D_2\,
     D_3\,D_4\, } \bigg[ ( -\alpha_1 + {\bar \alpha}_1 ) \,
        \Big( - \kkr\,\klklr  \\& \nonumber \quad + 
          \kkl\,\krklr \Big) \\& \nonumber + 
       \Big( \alpha_1\,\klkr + 
          {\bar \alpha}_1\,\kr^2 \Big) \,
        \Big( \frac{( -\alpha_1 + {\bar \alpha}_1 )}{\alpha_1} \,
             \kklr \\& \nonumber \quad- 
          \alpha_2\,\klr^2  + 
          \frac{{\alpha_2}^2\,{\bar \alpha}_1\, }{\alpha_1}
             ( \frac{\kr^2}{\alpha_1} + 
               \frac{ \bm{\ell}^2}{\alpha_{\ell}} )
           \Big) \\& \nonumber + \ep\,
        \bigg( \Big( - {\bar \alpha}_1\,\kklr   + 
             \alpha_2\,\klklr \Big) \,
           \Big( \frac{( \alpha_1 - {\bar \alpha}_2 ) \,}{\alpha_1}
                \kr^2 \\& \nonumber \quad + 2\,\krklr\
             \Big)  \\& \nonumber+ \Big( - {\bar \alpha}_1\,\kkr\
                 + \alpha_2\,\klkr \Big) \,
           \Big( - \alpha_2\,\klr^2  \\& \nonumber \quad - 
             2\,\krklr + 
             \alpha_1\,{\bar \alpha}_1\,
              ( \frac{\kr^2}{\alpha_1} + 
                \frac{ \bm{\ell}^2}{\alpha_{\ell}} )  \Big) 
          \bigg)  \bigg] \, \\& \nonumber+ 
  \frac{4\,\krklr}
   {\alpha_2\,{\bar \alpha}_1\,} \,\A\A_{23}^L \\[.3cm]
\A\A_{24}^T =&{}\; N_c\;\frac{1}{{D_2}^2\,D_4\,
       D_5\, }\,
       \bigg[ \frac{2\,\alpha_{\ell}\,( 1 - \ep ) \, }{{\bar \alpha}_1}
            \bigg( \alpha_1\,
               \Big( - {\bar \alpha}_1\,\kkl   + 
                 \alpha_2\,\kl^2 \Big) \,\kr^2 \\& \nonumber + 
              {\bar \alpha}_1\,\Big( - {\bar \alpha}_1\,
                    \kkr   + 
                 \alpha_2\,\klkr \Big) \,
               \Big( \kr^2 + \alpha_1\,Q^2\
                 \Big)  \bigg)\\& \nonumber + 
         \alpha_1\,\kr^2\,
          \Big( \alpha_{\ell}\,\kkl - 
            \alpha_2\,\kl^2 + 
            {\bar \alpha}_1\, \bm{k}(\bm{\ell} - 2\, \bm{r})
                 + \frac{\alpha_2\,{\bar \alpha}_1\,
                \bm{\ell}^2}{\alpha_{\ell}}\\& \nonumber \quad+ 
            \alpha_2\,( 2\,\bm{\ell} - \bm{r}) \, \kl 
                 \Big)  \\& \nonumber- 
         \alpha_2\,\Big( \kr^2 + 
            \alpha_1\,Q^2 \Big) \,
          \Big( \alpha_{\ell}\,\klkr + 
            {\bar \alpha}_1\,\kr^2 + 
            {\bar \alpha}_1\, \lpr \, \kr  
             \Big)  \bigg]  \\& \nonumber + 
  \frac{2\, \kr^2}
   {{\alpha_1}^2\,} \,\A\A_{24}^L\\[.3cm]
\A\A_{25}^T =&{}\;-\,N_c\;  \frac{2\,}{\alpha_1\,\alpha_{\ell}\,
    {D_2}^2\,D_4\,D_5}\alpha_2\,
    ( {\alpha_1}^2 + {\alpha_2}^2 ) \,{\bar \alpha}_1\,
    \kr^2 \, \bm{r}^2\\[.3cm]
\A\A_{33}^T =&{}\;  2\,\left[ \frac{ \klr^2  }
   {{{\bar \alpha}_1}^2\,} +
  \frac{ \bm{\ell}^2}
   {{\alpha_{\ell}}^2\,} \right]\,\A\A_{33}^L\\[.3cm]
\A\A_{34}^T =&{}\;- N_c\,  \frac{1}{D_1\,
     D_2\,D_3\,D_5\, }\bigg[ 2\,\ep\,
        \bigg( \frac{-2\,}{\alpha_2} {\bar \alpha}_1\,\alpha_{\ell}\,
             (\krklr)^2 \\& \nonumber + 
          \klkr\,\Big( - \alpha_2\,\alpha_{\ell}\,
                \klr^2   + 
             {\bar \alpha}_1\,\alpha_{\ell}\,\kr^2  + 
             \alpha_1\,{\bar \alpha}_1\, \bm{\ell}^2 \Big)  \\& \nonumber+ 
          \alpha_1\,{{\bar \alpha}_1}^2\, \bm{\ell}^2\,
           Q^2 + \frac{\alpha_{\ell}\, }{\alpha_1\,\alpha_2} \,\krklr\,
             \Big( {\bar \alpha}_1\,
                ( 2\,\alpha_1\,{\bar \alpha}_2 + 
                  \alpha_2\,
                   ( -{\bar \alpha}_1 + {\bar \alpha}_2 )  ) 
                 \,\kr^2 \\& \nonumber \quad + 
               2\,\alpha_1\,\alpha_2\,\lkl - 
               \alpha_1\,\alpha_2\,{{\bar \alpha}_1}^2\,
                Q^2 \Big) \\& \nonumber + 
          {\bar \alpha}_1\,\kr^2\,
           \Big( \frac{( \alpha_1 - {\bar \alpha}_2 ) \,
                \alpha_{\ell}\,\klklr}{\alpha_1\,
                {\bar \alpha}_1} + \alpha_{\ell}\,\klr^2 \\& \nonumber \quad+ 
             \frac{{\bar \alpha}_1\,\alpha_{\ell}\,\kr^2}
              {\alpha_1}  + {\bar \alpha}_1\, \bm{\ell}^2 + 
             {\bar \alpha}_1\,\alpha_{\ell}\,Q^2 \Big)  \bigg) \\& \nonumber
         + \frac{3\,( \alpha_1 - {\bar \alpha}_1 ) \,
          {\bar \alpha}_1\,\kr^2\,\rklr}{\alpha_1}\
        \\& \nonumber+ ( ( \alpha_1 - {\bar \alpha}_1 ) \,
           \klkr + \frac{{{\bar \alpha}_1}^2\,\kr^2}
           {\alpha_1} + \alpha_1\,{\bar \alpha}_1\,
           {\bar \alpha}_2\,Q^2 ) \,
        \lpr \, \klr  \\& \nonumber - 
       \frac{1 }
          {{\alpha_1}^2\,\alpha_{\ell}}\Big( {\alpha_1}^2\,\alpha_{\ell}\,\klr^2 - 
            \alpha_2\,{\bar \alpha}_1\,\alpha_{\ell}\,
             \kr^2 - \alpha_1\,\alpha_2\,
             {\bar \alpha}_1\, \bm{\ell}^2 \Big) \,\\& \nonumber \quad \cdot
          \Big( 2\,\alpha_1\,\alpha_{\ell}\,\klkr + 
            {\bar \alpha}_1\,\alpha_{\ell}\,\kr^2 + 
            \alpha_1\,{\bar \alpha}_1\,
              \lmrr \, \kr  \Big) \\& \nonumber - 
       \klklr\,\Big( \alpha_{\ell}\,\kr^2 + 
          \alpha_1\,{\bar \alpha}_2\,\alpha_{\ell}\,Q^2 + 
          ( \alpha_1 - {\bar \alpha}_1 ) \,
            \lpr \, \kr  \Big)\\& \nonumber  + 
       \krklr\,\Big( \frac{{\bar \alpha}_1\,}{\alpha_1\,\alpha_2}
             ( {\alpha_2}^2\,{\bar \alpha}_1 - 
               \alpha_1\,\alpha_2\,{\bar \alpha}_2 + 
               2\,\alpha_2\,
                ( {\bar \alpha}_1 - {\bar \alpha}_2 ) \,
                \alpha_{\ell} \\& \nonumber \quad- 
               4\,\alpha_1\,{\bar \alpha}_2\,\alpha_{\ell} ) \,
             \kr^2  - 
          2\,{\bar \alpha}_1\,{\bar \alpha}_2\,\alpha_{\ell}\,
           Q^2  \\& \nonumber \quad- \frac{( -\alpha_1 + 
               {\bar \alpha}_1 ) \, }{\alpha_{\ell}}
             ( - \alpha_{\ell}\,\kl^2   + 
               {\bar \alpha}_1\, \bm{\ell}^2 - 
               {\bar \alpha}_1\,\alpha_{\ell}\,Q^2 ) \\& \nonumber \quad +
               \frac{( -\alpha_1 + {\bar \alpha}_1\
               ) }
             {\alpha_2}( 2\,\alpha_{\ell}\,\klkr - 
               \alpha_2\, \lpr \, \kl\
                    \\& \nonumber \quad- 2\,{\bar \alpha}_1\,
                 \lpr \, \kr   ) \Big)  \bigg]  + 
  \frac{4\, \krklr}
   {\alpha_2\,{\bar \alpha}_1\,} \,\A\A_{34}^L\\[.3cm]
\A\A_{35}^T =&{}\;-\,N_c\;  \frac{ 2\,
    }{\alpha_{\ell}\,D_1\,D_2\,
    D_3\,D_5}\, {\bar \alpha}_1\,
    ( \alpha_2\,{\bar \alpha}_1 + 
      \alpha_1\,{\bar \alpha}_2 ) \,\krklr \, \bm{r}^2\\[.3cm]
\A\A_{44}^T =&{}\; N_c\;  \frac{2 \,}{{D_2}^2\,{D_5}^2\,
      }
     \bigg[ \frac{2\,{\alpha_{\ell}}^2\, }{
          \alpha_1\,{\bar \alpha}_1}( 1 - \ep ) \,
          \bigg( {\alpha_1}^2\,\kl^2\,\kr^2 + 
            {{\bar \alpha}_1}^2\,\kr^4 \\& \nonumber \quad + 
            {\alpha_1}^2\,{\bar \alpha}_1\,\klkr\,
             Q^2 \\& \nonumber \quad+ \alpha_1\,{\bar \alpha}_1\,
             (  2\,\kr^2 + \alpha_1\,Q^2\
               ) \,(  \klkr + 
               {\bar \alpha}_1\,Q^2 )  \bigg) \\& \nonumber  + 
       \alpha_1\,\Big( - \alpha_{\ell}\,\kl^2\,
             \kr^2   + 
          2\,{\bar \alpha}_1\,\kr^2\,
           \Big( \lmr^2 + \, \bm{r}^2 \Big)
               \\& \nonumber \quad + \alpha_{\ell}\,\kr^2\,
           \Big( - \frac{{\bar \alpha}_1\,\kr^2}
                {\alpha_1}   - {\bar \alpha}_1\,Q^2 + 
             3\, \lmr \,\kl   \Big) \
          \Big)  \\& \nonumber  - 3\,\alpha_{\ell}\,
        \Big( \kr^2 + \alpha_1\,Q^2 \Big) \,
        \Big( \alpha_{\ell}\,\klkr + 
          {\bar \alpha}_1\,\lmr\, \kr 
          \Big)  \bigg] \\& \nonumber + \frac{2\, \kr^2}
   {{\alpha_1}^2\,} \,\A\A_{44}^L \\[.3cm]
\A\A_{45}^T =&{}\;- N_c\;  \frac{ 2\,}{\alpha_1\,
    {D_2}^2\,{D_5}^2}( {\alpha_1}^2 + {\alpha_2}^2 ) \,
    {\bar \alpha}_1\,\kr^2 \, \bm{r}^2\\[.3cm]
\A\A_{55}^T =&{}\;0\\[.3cm]
\A\B_{11}^T =&{}\; - \frac{1}{N_c}\; \frac{2}{D_1\,D_2\,D_6\,
    D_7\, } \bigg[ \Big( {\bar \alpha}_1\,\kkr + 
         \alpha_1\,\krklr \Big) \,\\& \nonumber \quad \cdot
       \Big( \kl^2 - 2\,{\bar \alpha}_1\,\klkr + 
         \alpha_2\,{\bar \alpha}_1\,Q^2 \Big) \\& \nonumber + 
      \Big( {\bar \alpha}_1\,\kkl + 
         \alpha_1\,\klklr \Big) \,
       \Big( -2\,\alpha_1\,\klkr \\& \nonumber \quad+ \kr^2 + 
         \alpha_1\,{\bar \alpha}_2\,Q^2 \Big) \\& \nonumber + 
      \ep\,\bigg( \klkr\,
          \Big( ( \alpha_1\,{\bar \alpha}_1 + 
               \alpha_2\,{\bar \alpha}_2 ) \,\kklr - 
            2\,{\bar \alpha}_1\,{\bar \alpha}_2\,\kkr \\& \nonumber \quad - 
            2\,\alpha_1\,\alpha_2\,\klklr + 
            2\,\alpha_1\,{\bar \alpha}_1\,\klkr \Big) \\& \nonumber + 
         ( - \alpha_1\,{\bar \alpha}_1   + 
            \alpha_2\,{\bar \alpha}_2 ) \,
          \Big( \kkr\,\klklr \\& \nonumber \quad- 
            \kkl\,\krklr \Big)  \\& \nonumber + 
         {\bar \alpha}_1\,\kr^2\,
          \Big( \kkl + 
            \alpha_1\,{\bar \alpha}_1\,Q^2 \Big)  \\& \nonumber + 
         \alpha_1\,\kl^2\,
          \Big( \krklr  + 
            \alpha_1\,{\bar \alpha}_1\,Q^2 \Big)  \\& \nonumber + 
         \alpha_1\,{\bar \alpha}_1\,Q^2\,
          \Big( {\bar \alpha}_1\,\kkl +
            \alpha_1\,\krklr + 
            \alpha_1\,{\bar \alpha}_1\,Q^2 \Big)  \bigg) \
      \bigg]\\[.3cm]
\A\B_{12}^T =&{}\;- C_F\;  \frac{4 \,}{D_1\,D_2\,
    D_3\,D_7\, }
    \bigg[ \frac{1}{\alpha_1}( \alpha_2\,{{\bar \alpha}_1}^2 - 
           2\,\alpha_1\,{\bar \alpha}_1\,{\bar \alpha}_2 - 
           \alpha_1\,{{\bar \alpha}_2}^2 ) \,\klklr\,
         \kr^2 \\& \nonumber - 
      \alpha_1\,{{\bar \alpha}_2}^2\,\klklr\,
       Q^2 \\& \nonumber + \frac{1 }{\alpha_1}\klkr\,
         \Big( \alpha_1\,
            ( \alpha_2\,{\bar \alpha}_1 + 
              \alpha_1\,{\bar \alpha}_2 ) \,\klr^2 \\& \nonumber \quad+ 
           {\bar \alpha}_1\,( {\bar \alpha}_1 - {\bar \alpha}_2\
              ) \,{\bar \alpha}_2\,\kr^2 + 
           \alpha_1\,{\bar \alpha}_1\,{\bar \alpha}_2\,
            ( 4\,\krklr - {\bar \alpha}_2\,Q^2\
              )  \Big)\\& \nonumber  + 
      \ep\,\bigg( \frac{( \alpha_1\,
               {{\bar \alpha}_1}^2 + 
              2\,\alpha_1\,{\bar \alpha}_1\,{\bar \alpha}_2 - 
              \alpha_2\,{{\bar \alpha}_2}^2 )}{\alpha_1} \,
            \klklr\,\kr^2 \\& \nonumber+ 
         \alpha_1\,{{\bar \alpha}_1}^2\,\klklr\,
          Q^2 \\& \nonumber + \frac{1 }{\alpha_1}\klkr\,
            \Big( \alpha_1\,
               ( \alpha_1\,{\bar \alpha}_1 + 
                 \alpha_2\,{\bar \alpha}_2 ) \,\klr^2 \\& \nonumber \quad- 
              {\bar \alpha}_1\,( {\bar \alpha}_1 - {\bar \alpha}_2\
                 ) \,{\bar \alpha}_2\,\kr^2  + 
              \alpha_1\,{\bar \alpha}_1\,{\bar \alpha}_2\,
               ( -4\,\krklr - 
                 {\bar \alpha}_1\,Q^2 )  \Big)
            \bigg)  \bigg]\\[.3cm]
\A\B_{13}^T =&{}\; \frac{1}{N_c}\;  \frac{2 \,}{D_1\,
     D_2\,D_4\,D_6
      }\bigg[ \Big( {\bar \alpha}_1\, \bm{k}^2 - 
          \alpha_2\,\kklr \Big) \,
        \Big( \alpha_2\,\klkr - 
          {\bar \alpha}_1\,\kr^2 \Big) \\& \nonumber + 
       ( {\alpha_1}^2 - {\alpha_2}^2 ) \,
        \Big( \kkr\,\klklr - 
          \kkl\,\krklr \Big) \\& \nonumber - 
       \alpha_1\,\Big( -\kklr + \kl^2 + 
          \frac{{\bar \alpha}_1\,}{\alpha_{\ell}}\, \bm{\ell}^2 \Big) \\& \nonumber \quad\,
        \cdot \,\Big( \alpha_1\,\klkr + 
          {\bar \alpha}_1\,\kr^2 + 
          \alpha_1\,{\bar \alpha}_1\,Q^2 \Big) \\& \nonumber + 
       \ep\,\bigg( - {\bar \alpha}_1\,
             \Big( - {\bar \alpha}_1\, \bm{k}^2   + 
               \alpha_2\,\kkl \Big) \,\kr^2\
              \\& \nonumber \quad+ ( {\alpha_1}^2 - {\alpha_2}^2\
             ) \,\Big( - \kkr\,\klklr\
                 \\& \nonumber \qquad+ \kklr\,\klkr  - 
             \kkl\,\krklr \Big) \\& \nonumber \quad + 
          \krklr\,\Big( - \alpha_2\,{\bar \alpha}_1\,
                 \bm{k}^2   + 
             {\alpha_1}^2\,\kl^2 + 
             \frac{{\alpha_1}^2\,{\bar \alpha}_1}
              {\alpha_{\ell}}\, \bm{\ell}^2\, \Big) \\& \nonumber\quad  + 
          \frac{\alpha_1\,{\bar \alpha}_1\, }
             {\alpha_{\ell}}
             \Big( {\bar \alpha}_1\, \bm{\ell}^2 + 
               \alpha_{\ell}\,\lkl \Big) \,
             \Big( \kr^2 + \alpha_1\,Q^2 \Big) \bigg)  \bigg] \\& \nonumber + \frac{4\, \kkr}
   {{\alpha_1}^2\,} \,\A\B_{13}^L\\[.3cm]
\A\B_{14}^T =&{}\; N_c\; \frac{1}{D_1\,D_2\,
       D_5\,D_7\, }
       \bigg[ -2\,\ep\,
          \bigg( \frac{-2\,}{\alpha_2} {\bar \alpha}_1\,\alpha_{\ell}\,
               [\klkr]^2 \\& \nonumber+ 
            \frac{( \alpha_1\,{\bar \alpha}_1 - 
                 \alpha_2\,{\bar \alpha}_2 ) \,\alpha_{\ell}\,}{\alpha_1}
               \klklr\,\kr^2 \\& \nonumber - 
            \alpha_{\ell}\,\krklr\,
             \Big( \kl^2 + 
               \alpha_1\,{\bar \alpha}_1\,Q^2 \Big)  \\& \nonumber - 
            {\bar \alpha}_1\,\alpha_{\ell}\,Q^2\,
             \Big(  \alpha_1\,\kl^2   +
               {\bar \alpha}_1\,\kr^2 + 
               \alpha_1\,{\bar \alpha}_1\,Q^2 \Big)\\& \nonumber  + 
            \alpha_{\ell}\,\klkr\,
             \Big( 2\,\lmr \, \klr  \\& \nonumber \quad+ 
               ( \frac{2\,{\bar \alpha}_1\,{\bar \alpha}_2}
                   {\alpha_2} + 
                  \frac{{\bar \alpha}_1\,
                     ( -{\bar \alpha}_1 + {\bar \alpha}_2 ) }{\alpha_1}
                     ) \,\kr^2  
                - {{\bar \alpha}_1}^2\,Q^2\
               \Big)  \bigg) \\& \nonumber -
         ( \kr^2 + 
            \alpha_1\,{\bar \alpha}_2\,Q^2 ) \,
          \Big( - \alpha_{\ell}\,\klklr   + 
            {\bar \alpha}_1\, \lmrr \,\kl
                 \Big)  \\& \nonumber - 
         \frac{1 }{\alpha_1}\kr^2\,
            \Big( - {\bar \alpha}_1\,\alpha_{\ell}\,
                 \kl^2   - 
              \alpha_2\,{{\bar \alpha}_1}^2\,\alpha_{\ell}\,
               Q^2 + 
              {\bar \alpha}_1\,( -\alpha_1 + {\bar \alpha}_1\
                 ) \, \lpr \, \kl  \
              \Big) \\& \nonumber+ 
         \klkr\,\Big[ - \frac{{\bar \alpha}_1\,
                 ( -\alpha_1 + {\bar \alpha}_1 ) \,
                 ( \alpha_2 + 2\,\alpha_{\ell} ) \,}{\alpha_1\,\alpha_2}
                 \kr^2   \\& \nonumber \quad+ 
            \frac{2\,{\bar \alpha}_1\,{\bar \alpha}_2\,\alpha_{\ell}\, }{\alpha_2}
               ( \frac{1}{\alpha_1}\kr^2 + 
                 \alpha_2\,Q^2 ) \\& \nonumber \quad + 
            \frac{( -\alpha_1 + {\bar \alpha}_1 ) \, }{\alpha_2\,\alpha_{\ell}}
               \Big(\, \alpha_2\,
                  ( - \alpha_{\ell}\,\kl^2   + 
                    {\bar \alpha}_1\, \bm{\ell}^2 - 
                    {\bar \alpha}_1\,\alpha_{\ell}\,Q^2 + 
                    \alpha_{\ell}\,
                      \lmrr \,\klr  
                    ) \\& \nonumber\qquad - 2\,\alpha_{\ell}\,
                  ( \alpha_{\ell}\,\krklr - 
                    {\bar \alpha}_1\,
                      \lmrr \,\kr  
                    )  \Big)
            \Big] \\& \nonumber + ( -\alpha_1 + {\bar \alpha}_1 ) \,
          \Big( \krklr\,
              \lmrr \,\kl  \\& \nonumber \quad- 
            \klklr\, \lmrr 
               \,\kr   \Big)\\& \nonumber  + 
         ( \kl^2 + 
            \alpha_2\,{\bar \alpha}_1\,Q^2 ) \,
          \Big( 2\,\alpha_{\ell}\,\krklr + 
            {\bar \alpha}_1\,\lpr \,\kr  
             \Big)  \bigg]  \\& \nonumber + 
  \frac{4\, \klkr}
   {\alpha_2\,{\bar \alpha}_1\,}\,\A\B_{14}^L\\[.3cm]
\A\B_{15}^T =&{}\;- N_c\;  \frac{2\,
    }{\alpha_{\ell}\,D_1\,D_2\,
    D_5\,D_7}\,{\bar \alpha}_1\,
    ( \alpha_2\,{\bar \alpha}_1 + 
      \alpha_1\,{\bar \alpha}_2 ) \,\klkr\,\bm{r}^2\\[.3cm]
\A\B_{22}^T =&{}\;- \frac{1}{N_c}\;  \frac{2\,}{
    D_2\,D_3\,D_4\,D_7\,
     }\bigg[ \frac{( \alpha_1 - {\bar \alpha}_1\
           )}{\alpha_1} \,\kr^2\,\Big( {\bar \alpha}_2\,\kkl + 
           \alpha_2\,\klklr \Big)  \\& \nonumber-
         \frac{( - \alpha_2\,
              {\bar \alpha}_1   + \alpha_1\,{\bar \alpha}_2\
           ) }{{\bar \alpha}_1} \,\kl^2\,
         \Big( {\bar \alpha}_2\,\kkr + 
           \alpha_2\,\krklr \Big) \\& \nonumber - 
      2\,\klkr\,\Big( {{\bar \alpha}_2}^2\,\kkr + 
         {\alpha_2}^2\,\klklr \\& \nonumber \quad + 
         \alpha_2\,{\bar \alpha}_2\,
          ( \kkl + \krklr )  \Big) \\& \nonumber + 
      \ep\,\bigg(  \klkr\, \Big( ( \alpha_1\,{\bar \alpha}_1 + 
               \alpha_2\,{\bar \alpha}_2 ) \,\kklr - 
            2\,{\bar \alpha}_1\,{\bar \alpha}_2\,\kkr  \\& \nonumber \quad  - 
            2\,\alpha_1\,\alpha_2\,\klklr \Big) \,
          + 2\,\alpha_2\,{\bar \alpha}_2\,
          [\klkr]^2 \\& \nonumber + 
         ( \alpha_1\,{\bar \alpha}_1 - 
            \alpha_2\,{\bar \alpha}_2 ) \,
          \Big( \kkr\,\klklr - 
            \kkl\,\krklr \Big) \\& \nonumber + 
         ( \alpha_1\,{\bar \alpha}_1 - 
            \alpha_2\,{\bar \alpha}_2 ) \,
          \Big( \frac{{\bar \alpha}_2}
             {\alpha_1}\,\kkl\,\kr^2 - 
            \frac{\alpha_2\,{\bar \alpha}_2}{\alpha_1\,{\bar \alpha}_1}\,\kl^2\,
               \kr^2 \\& \nonumber \quad + 
            \frac{\alpha_2}
             {{\bar \alpha}_1}\,\kl^2\,\krklr \Big)  \bigg)  \bigg]\\[.3cm]
\A\B_{23}^T =&{}\;- C_F\;  \frac{4 \,
     }{{\bar \alpha}_1\,D_2\,{D_4}^2\,D_6\,
      }\,{\alpha_1}^2\,( 1 - \ep ) \,
     \bigg( {\bar \alpha}_1\, \bm{k}^2 - 
       \alpha_2\,\kl^2 - 
       \frac{\alpha_2\,{\bar \alpha}_1\, \bm{\ell}^2}{\alpha_{\ell}}\
       \bigg) \,\lkr\, \\& \nonumber +
  \frac{2\, \kkr}
   {{\alpha_1}^2\,} \,\A\B_{23}^L\\[.3cm]
\A\B_{24}^T =&{}\;- N_c\;  \frac{1}{D_2\,
     D_4\,D_5\,D_7\, }\, \bigg[ 2\,\ep\,
        \bigg[ \frac{\alpha_2\,
             ( - \alpha_1\,{\bar \alpha}_1   + 
               \alpha_2\,{\bar \alpha}_2 ) \,\alpha_{\ell}\,}{\alpha_1\,
             {\bar \alpha}_1}
             \kl^2\,\kr^2 \\& \nonumber + 
          \alpha_{\ell}\,\kkl\,
           \Big( 2\,\klkr + 
             \frac{( \alpha_1\,{\bar \alpha}_1 - 
                  \alpha_2\,{\bar \alpha}_2 )}
                {\alpha_1} \,\kr^2 \Big) \\& \nonumber - 
          \alpha_{\ell}\,\kkr\,
           \Big( \kl^2 + 2\,\klkr + 
             \alpha_1\,{\bar \alpha}_1\,Q^2 \Big)\\& \nonumber  - 
          \frac{\alpha_2\,\alpha_{\ell} } {{\bar \alpha}_1}\,\klkr\,
             \Big( \kl^2 - 2\,\klkr - 
               \alpha_1\,{\bar \alpha}_1\,Q^2 \Big)
             \bigg] \\& \nonumber + 
       \klkr\,\bigg( \frac{2\,\alpha_2\,\alpha_{\ell}\,
             ( \kl^2 +
               \alpha_2\,{\bar \alpha}_1\,Q^2 ) }{{\bar \alpha}_1} \\& \nonumber\quad+
            (  {\bar \alpha}_1 - \alpha_1 \
             ) \,\Big( \alpha_2\,
              ( - \frac{\kl^2}{{\bar \alpha}_1}
                    - \frac{\kr^2}{\alpha_1} + 
                \frac{ \bm{\ell}^2}{\alpha_{\ell}} - Q^2\
                )  \\& \nonumber \quad + \frac{1 }{{\bar \alpha}_1} (-2\,\alpha_{\ell}\,\kkl + 
                {\bar \alpha}_1\,
                 \bm{k} \, \lmrr  + 
                2\,\alpha_2\,
                  \lmrr\,\kl )
                \Big)  \bigg) \\& \nonumber + 
       (  {\bar \alpha}_1 \, - \alpha_1  ) \,
        \bigg( - \kkr\,
              \lmrr \,\kl +  \kkl\, \lmrr \,\kr   \\& \nonumber \quad - 
          \frac{1 }{\alpha_1}\kr^2\,
             \Big( 2\,\alpha_{\ell}\,\kkl + 
               \alpha_2\, \lpr \,\kl\
                    \Big) \\& \nonumber \quad + 
          \frac{1 }{{\bar \alpha}_1}\kl^2\,
             \Big( \alpha_{\ell}\,\kkr - 
               \alpha_2\, \lmrr \kr 
                    \Big) 
               \bigg) \\& \nonumber + 
       \frac{\alpha_2\,}{\alpha_1}\,\Big( \kl^2 + 
            \alpha_2\,{\bar \alpha}_1\,Q^2 \Big) \,
          \Big( \alpha_{\ell}\,\kr^2 + 
            \alpha_1\, \lpr \, \kr  
            \Big)  \bigg] \\& \nonumber + 
  \frac{4\, \klkr}
   {\alpha_2\,{\bar \alpha}_1\,} \,\A\B_{24}^L\\[.3cm]
\A\B_{25}^T =&{}\;N_c\;  \frac{2\,
    }{\alpha_{\ell}\,D_2\,D_4\,
    D_5\,D_7}\,\alpha_2\,
    ( \alpha_2\,{\bar \alpha}_1 + 
      \alpha_1\,{\bar \alpha}_2 ) \,\klkr \, \bm{r}^2\\[.3cm]
\A\B_{33}^T =&{}\;- \frac{1}{N_c}\;  \frac{2 \,}{
    D_1\,D_3\,D_4\,D_6\,
     }\bigg[ - \Big( \alpha_1\,\klklr + 
           {\bar \alpha}_1\,\krklr \Big) \\& \nonumber \quad \cdot
         \Big( - {\bar \alpha}_1\, \bm{k}^2   + 
           2\,\alpha_1\,\kklr + 
           \alpha_1\,{\bar \alpha}_2\,
            ( \frac{\kl^2}{{\bar \alpha}_1} + 
              \frac{ \bm{\ell}^2}{\alpha_{\ell}} )  \Big) \\& \nonumber  
        - \Big( \alpha_1\,\kkl + 
         {\bar \alpha}_1\,\kkr \Big) \,
       \Big( 2\,{\bar \alpha}_1\,\kklr - 
         \alpha_1\,\klr^2 \\& \nonumber \quad + 
         \alpha_2\,{\bar \alpha}_1\,
          ( \frac{\kr^2}{\alpha_1}  + 
            \frac{ \bm{\ell}^2}{\alpha_{\ell}} )  \Big)  \\& \nonumber + 
      \ep\,\bigg( ( \alpha_1\,{\bar \alpha}_1 + 
            \alpha_2\,{\bar \alpha}_2 ) \,\kklr\,
          \klkr  \\& \nonumber\quad+ {\bar \alpha}_2\,
           {\bar \alpha}_1\,
             \Big( -2\,\kklr\,\kkr + 
                \bm{k}^2\,\krklr \Big)  \\& \nonumber \quad+ 
             {\bar \alpha}_2\,\alpha_2\,\Big( \kkr\,\klklr - 
               \kkl\,\krklr \Big)  \\& \nonumber \quad  + 
         \alpha_1\, \alpha_2\,
             \Big( -2\,\kklr\,\klklr + 
               \kkl\,\klr^2 \Big)  \\& \nonumber \quad- 
             \alpha_1\,{\bar \alpha}_1\,\Big( \kkr\,\klklr - 
               \kkl\,\krklr \Big)    \\& \nonumber \quad+ 
         \frac{\alpha_1\,{{\bar \alpha}_1}^2\, }{\alpha_{\ell}}
            \Big( - \alpha_{\ell}\,\krklr   + 
              \alpha_1\, \bm{\ell}^2 \Big) \,
            \Big( \frac{\kl^2}{{\bar \alpha}_1} + 
              \frac{ \bm{\ell}^2}{\alpha_{\ell}} \Big)\\& \nonumber\quad
          + \frac{{\alpha_1}^2\,{\bar \alpha}_1\, }{\alpha_{\ell}}
            \Big( - \alpha_{\ell}\,\kkl   + 
              {\bar \alpha}_1\, \bm{\ell}^2 \Big) \,
            \Big( \frac{\kr^2}{\alpha_1} + 
              \frac{ \bm{\ell}^2}{\alpha_{\ell}} \Big)\\& \nonumber\quad
          - \alpha_1\,{\bar \alpha}_1\,
          \Big( -2\,(\kklr)^2 + 
             \bm{k}^2\,\klr^2 \\& \nonumber \qquad- 
            \kl^2\,\kr^2 + 
            \frac{\alpha_1\,{\bar \alpha}_1\,\bm{\ell}^4}
             {{\alpha_{\ell}}^2} \Big)  \bigg)  \bigg]\\[.3cm]
\A\B_{34}^T =&{}\;- N_c\;  \frac{1}{D_1\,D_3\,
     D_5\,D_7\, }\bigg[ 2\,\ep\,
        \bigg( \alpha_{\ell}\,\Big( \alpha_1\,\kl^2 - 
             {\bar \alpha}_2\,\klkr \Big) \,\klr^2 \\& \nonumber\quad - 
          \frac{( {\bar \alpha}_1 - {\bar \alpha}_2 ) \,
             \alpha_{\ell}}{\alpha_1}\, \klklr\,\kr^2 \\& \nonumber\quad-
             \alpha_{\ell}\,\krklr\,
           \Big( \kl^2 + {{\bar \alpha}_1}^2\,Q^2\
             \Big)  \\& \nonumber\quad + {\bar \alpha}_1\,\alpha_{\ell}\,
           \Big( \frac{\kr^2}{\alpha_1} + 
             \frac{ \bm{\ell}^2}{\alpha_{\ell}} \Big) \,
           \Big( {\bar \alpha}_1\,\klkr + 
             \alpha_1\,( \kl^2 + 
                {\bar \alpha}_1\,Q^2 )  \Big)  \bigg)  \\& \nonumber - 
       {\bar \alpha}_1\,\Big( \frac{\kr^2}{\alpha_1} + 
          \frac{ \bm{\ell}^2}{\alpha_{\ell}} \Big) \,
        \Big( {\bar \alpha}_1\,\alpha_{\ell}\,\klkr \\& \nonumber \quad+ 
          2\,\alpha_1\,\alpha_{\ell}\,
           \Big( \kl^2 + {\bar \alpha}_1\,Q^2 \Big) 
          + \alpha_1\,{\bar \alpha}_1\,
            \lmrr \,\kl   \Big) \\& \nonumber + 
       \klr^2\,\Big( -2\,\alpha_1\,\alpha_{\ell}\,
           \kl^2 + {\bar \alpha}_2\,\alpha_{\ell}\,
           \klkr  \\& \nonumber\quad+ \alpha_1\,{\bar \alpha}_2\,
           ( \lmrr \,\kl )  \Big) \\& \nonumber + 
       3\,( {\bar \alpha}_1 - {\bar \alpha}_2 ) \,
        \Big( - \krklr\,\rkl   + 
          \klkr\,\rklr \Big) \\& \nonumber + 
       ( \kl^2 + {{\bar \alpha}_1}^2\,Q^2 ) \,
        \Big( \alpha_{\ell}\,\krklr + 
          \alpha_1\, \lpr \, \klr 
          \Big)  \\& \nonumber + \frac{( {\bar \alpha}_1 - {\bar \alpha}_2 ) \,
         }{\alpha_1\,{\bar \alpha}_1\,
          \alpha_{\ell}}
          \klklr\,\bigg( 2\,{\bar \alpha}_1\,{\alpha_{\ell}}^2\,
             \kr^2 + {\alpha_1}^2\,{\bar \alpha}_1\,
              \bm{\ell}^2 \\& \nonumber \quad- {\alpha_1}^2\,\alpha_{\ell}\,
             \Big( \kl^2 + {\bar \alpha}_1\,Q^2 \Big) 
              - \alpha_1\,{\bar \alpha}_1\,\alpha_{\ell}\,
             \Big( \klkr + \lkr\
               \Big)  \bigg) \bigg]  \\& \nonumber+ 
  \frac{4\, \klklr}
   {{{\bar \alpha}_1}^2\,}\,\A\B_{34}^L\\[.3cm]
\A\B_{35}^T =&{}\;N_c\;  \frac{2\,}{\alpha_{\ell}\,D_1\,
    D_3\,D_5\,D_7}\alpha_1\,
    ( {{\bar \alpha}_1}^2 + {{\bar \alpha}_2}^2 ) \,
    \klklr \, \bm{r}^2\\[.3cm]
\A\B_{44}^T =&{}\;- N_c\;  \frac{1}{D_2\,{D_5}^2\,D_7\,
     }\bigg[ -4\,{\alpha_{\ell}}^2\,
       (\klkr)^2 + \frac{( \alpha_1 + 
           {\bar \alpha}_1 ) \,{\alpha_{\ell}}^2\,}{\alpha_1\,{\bar \alpha}_1}\,\kl^2\,
         \kr^2 \\ &+ 
      4\,\ep\,\bigg( 4\,{\alpha_{\ell}}^2\,(\klkr)^2 - 
         \frac{( \alpha_1 - {\bar \alpha}_2 ) \,
            {\alpha_{\ell}}^2}{
            \alpha_1\,{\bar \alpha}_1}\,\kl^2\,\kr^2 \\& \nonumber\quad +
         {\alpha_{\ell}}^2\,Q^2\,
          \Big( \alpha_1\,\kl^2 + 
            {\bar \alpha}_1\,\kr^2  + 
            \alpha_1\,{\bar \alpha}_1\,Q^2 \Big)  \\& \nonumber\qquad - 
         \frac{{\alpha_{\ell}}^2 }{\alpha_1\,{\bar \alpha}_1}\,\klkr\,
            \Big( \alpha_1\,
               ( -\alpha_1 + \alpha_2 ) \,
               \kl^2 + 
              {\bar \alpha}_1\,( -{\bar \alpha}_1 + {\bar \alpha}_2\
                 ) \,\kr^2 \\& \nonumber \quad -
              \alpha_1\,{\bar \alpha}_1\,
               ( \alpha_1 + {\bar \alpha}_1 ) \,
               Q^2 \Big)
         \bigg) \\& \nonumber + \frac{{\bar \alpha}_1\,\alpha_{\ell} }{\alpha_1}\,\kr^2\,
         \Big(  \alpha_2\,\alpha_{\ell}\,Q^2\
                - 3\, \lmr \, \kl  \Big) \\& \nonumber + 
      \frac{\alpha_1\,\alpha_{\ell} }{{\bar \alpha}_1}\,\kl^2\,
         \Big(  {\bar \alpha}_2\,\alpha_{\ell}\,Q^2\
                + 3\, \lmr \, \kr  \
           \Big) \\& \nonumber + 
      \klkr\,\bigg( \frac{-2\,\alpha_{\ell}\,
            ( \alpha_2\,{\bar \alpha}_1 + 
              \alpha_1\,{\bar \alpha}_2 + 
              2\,( \alpha_1 - \alpha_2 ) \,
               \alpha_{\ell} )}{{\bar \alpha}_1} \,\kl^2 \\& \nonumber \quad- 
         \frac{2\,\alpha_{\ell}\,
            ( \alpha_2\,{\bar \alpha}_1 + 
              \alpha_1\,{\bar \alpha}_2 + 
              2\,( {\bar \alpha}_1 - {\bar \alpha}_2 ) \,
               \alpha_{\ell} ) }{\alpha_1}\,\kr^2 \\& \nonumber\quad + 
         4\,( \alpha_2 + {\bar \alpha}_2 ) \,
          {\alpha_{\ell}}^2\,Q^2 + 
         2\,( \alpha_2\,{\bar \alpha}_1 + 
            \alpha_1\,{\bar \alpha}_2 ) \,
          \Big(- \alpha_{\ell}\,Q^2 + 2\, \bm{r}^2 + 
            2\,\lmr^2
                 \Big)  \\& \nonumber \quad- 
         6\,( \alpha_1 - \alpha_2 ) \,\alpha_{\ell}\,
          \lmr \,\kl   + 
         6\,( {\bar \alpha}_1 - {\bar \alpha}_2 ) \,
          \alpha_{\ell}\, \lmr \, \kr  \
         \bigg)  \\& \nonumber + 3\,\alpha_2\,{\bar \alpha}_1\,\alpha_{\ell}\,
       Q^2\, \lmr \, \kr - 3\,\alpha_1\,{\bar \alpha}_2\,\alpha_{\ell}\,Q^2\,    \lmr \, \kl
      \bigg]\\[.3cm]
\A\B_{45}^T =&{}\;N_c\;  \frac{2\,}{D_2\,{D_5}^2\,D_7}\,
        ( \alpha_2\,{\bar \alpha}_1 + 
      \alpha_1\,{\bar \alpha}_2 ) \,\klkr \, \bm{r}^2\\[.3cm] 
\A\B_{55}^T =&{}\;0
\end{align}

At several places we have expressed the matrix elements $\A\A_{(ij)}^T$,
$\A\B_{(ij)}^T$ in terms of the corresponding matrix elements for the 
longitudinally polarized photon; we therefore list here also the expressions 
for longitudinal polarizations \cite{BGK}:
\begin{align}
\A\A_{11}^L=&{}\;
   C_F \;\frac{2}{D_1^2\,D_2^2}\,(1-\epsilon)\,\alpha_1\,\bar{ \alpha }_1\,\bm{A}_1^2
      \label{eq:AA11}\\
\A\A_{12}^L=&{}\; 
   - \frac{1}{N_c} \;\frac{1}{D_1\,D_2^2\,D_4}\left(\,(1-\epsilon)\,\alpha_1\,
        \alpha_2 \,  \bm{A}_1 \,\bm{A}_2  - 
     \alpha_1\,\alpha_2^2\,\bar{ \alpha }_1\,\bm{r}^2 \right) \\
\A\A_{13}^L=&{}\;
  C_F\;\frac{2}{D_1^2 \,D_2\,D_3}\,\left( (1-\epsilon)\,
        \bar{ \alpha }_1^2 \, \bm{A}_1\,\bm{A}_3\, + \bar{ \alpha }_1^2\,\left( \bm{k} + \bm{r} \right) \,
        \left( \bm{k} + \bm{\ell} + \bm{r} \right)  \right)  \\
\A\A_{14}^L=&{}\;
   - N_c \; \frac{1} {D_1\,D_2^2\,D_5}\,\bigg( \frac{1}{2\,
          \alpha_\ell}\,
            \alpha_1\,\left( \alpha_2\,\bar{ \alpha }_1 + 2\,(1-\epsilon)\,\alpha_\ell^2 
      \right)  \bm{A}_1\,\bm{A}_4
   \nonumber \\* 
   &+ \frac{1}{2}\alpha_1\,\alpha_2\,
          \bar{ \alpha }_1^2\,D_2 + 
       \frac{1}{2\,\alpha_\ell}\,\alpha_1\,\alpha_2^2\,
          \bar{ \alpha }_1\, \bm{A}_1 \bm{r}- 
       \alpha_1\,\alpha_2^2\,\bar{ \alpha }_1\,\bm{r}^2 \bigg)
     \\
\A\A_{15}^L=&{}\;
   N_c\;\frac{1}{\alpha_\ell\,D_1\,D_2^2\,D_5}\; \alpha_1\,\alpha_2^2\,\bar{ \alpha }_1^2\,
    \bm{r}^2 \\
\A\A_{22}^L=&{}\;
   C_F\; \frac{2}{\bar{ \alpha }_1\,D_2^2\,D_4^2}\;(1-\epsilon)\,\alpha_1\,\alpha_2^2\,\bm{A}_2^2 \\
\A\A_{23}^L=&{}\;
   - \frac{1}{N_c} \;\frac{1} {D_1\,D_2\,D_3\,D_4}\left( (1-\epsilon)\,\alpha_2\,
        \bar{ \alpha }_1 \,  \bm{A}_2\,\bm{A}_3\, + 
     \alpha_2\,\bar{ \alpha }_1\,
      \left( \bm{k} + \bm{r} - \alpha_1\, \bm{r} \right) \,
      \left( \bm{k} + \bm{\ell} + \bar{ \alpha }_1\, \bm{r} \right) \right)\\
\A\A_{24}^L=&{}\;
  N_c\;\frac{1}{D_2^2\,D_4\,D_5}\bigg(
  \frac{1}{2\,\alpha_\ell\,\bar{ \alpha }_1} \, 
          \alpha_1\,\alpha_2\,
          \left( \alpha_2\,\bar{ \alpha }_1 + 2\,(1-\epsilon)\,\alpha_\ell^2
            \right)  \bm{A}_2\,\bm{A}_4\, \nonumber \\*
     &+ \frac{1}{2} \alpha_1\,\alpha_2^2\,\bar{ \alpha }_1\,
          D_2   + 
       \frac{1}{2\,\alpha_\ell}\,\alpha_1\,\alpha_2^2\,
          \bar{ \alpha }_1\, \bm{A}_2 \bm{r} + 
       \alpha_1\,\alpha_2^2\,\bar{ \alpha }_1\,\bm{r}^2 \bigg) 
      \\
\A\A_{25}^L=&{}\;
   - N_c\; \frac{1}{\alpha_\ell\,D_2^2\,D_4\,D_5}\; \alpha_1\,\alpha_2^3\,\bar{ \alpha }_1\,\bm{r}^2 \\
\A\A_{33}^L=&{}\;
  C_F \;\frac{2}{\alpha_1\,D_1^2\,D_3^2}\;(1-\epsilon)\,\bar{ \alpha
  }_1^3\,\bm{A}_3^2 \\
\A\A_{34}^L=&{}\;
  N_c \;\frac{1}{D_1\,D_2\,D_3\,D_5}\,\bigg( \frac{1}{2\,\alpha_\ell}
          \bar{ \alpha }_1\,\left( \alpha_1\,\bar{ \alpha
  }_1-\alpha_\ell - 2\,(1-\epsilon)\,\alpha_\ell^2 \right)\,\bm{A}_3\,\bm{A}_4\, 
   \nonumber \\* 
    &- \frac{1}{2}\, \alpha_1\,\bar{ \alpha }_1^2\,\bar{ \alpha }_2\,
          D_2 + 
       \alpha_1\,\alpha_2\,\bar{ \alpha }_1^2\,
        D_7  + 
       \frac{1}{2\,\alpha_\ell}\,\alpha_1\,\alpha_2\,
          \bar{ \alpha }_1^2\, \bm{A}_3 \bm{r}+
  \alpha_1\,\alpha_2\,\bar{ \alpha }_1^2\,\bm{r}^2 \nonumber \\* 
  &+ \frac{3}{2}
  \,\alpha_2\, \bar{ \alpha }_1\, \bm{A}_3 \left( \bm{k} + \bm{\ell} \right)
\bigg) \\
\A\A_{35}^L=&\,
   - N_c\; \frac{1}{\alpha_\ell\,D_1\,D_2\,D_3\,D_5}\,\alpha_1\,\alpha_2\,\bar{ \alpha }_1^2\,
       \bar{ \alpha }_2\,\bm{r}^2 \\
\A\A_{44}^L=&{}\;
   N_c \;\frac{1}{D_2^2\,D_5^2}\, \left( \frac{2}{\bar{ \alpha }_1}\,\bm{A}_4^2\,\alpha_1\,
            \left( (1-\epsilon)\,\alpha_\ell^2 + \alpha_2\,\bar{ \alpha }_1 \right)
               + 
         \alpha_1\,\alpha_2\,\alpha_\ell\,\bar{ \alpha }_1\,
          D_2 + 2\,\alpha_1\,\alpha_2^2\,
          \bar{ \alpha }_1\,\bm{r}^2 \right)   \\
\A\A_{45}^L=&{}\;
   - N_c \;\frac{1}{D_2^2\,D_5^2} \alpha_1\,\alpha_2^2\,\bar{ \alpha }_1\,
       \bm{r}^2 \\
\A\A_{55}^L=&\;0 \\
\A\B_{11}^L=&{}\;
 \frac{1}{N_c}\; \frac{1}{D_1\,D_2\,D_6\,D_7} \left((1-\epsilon)\,\alpha_1\,\bar{
  \alpha }_1 \,  \bm{A}_1\,\bm{B}_1 + 
     \alpha_1\,\bar{ \alpha }_1\,
      \left( \bm{k} + \alpha_1\,\bm{r} \right) \,
      \left( \bm{k} + \bm{\ell} + \bar{ \alpha }_2\,\bm{r} \right) \right) \\
\A\B_{12}^L=&{}\;
   - C_F \;\frac{2}{D_1\,D_2\,D_3\,D_7}\,\left(  \,(1-\epsilon)\,\bar{ \alpha }_1\,
          \bar{ \alpha }_2\,\bm{A}_1 \,\bm{B}_2   + 
       \bar{ \alpha }_1\,\bar{ \alpha }_2\,\left( \bm{k} + \bm{r} \right) \,
        \left( \bm{k} + \bm{\ell} + \bm{r} \right)  \right)  \\
\A\B_{13}^L=&{}\;
  \frac{1}{N_c}\frac{1} {D_1\,D_2\,D_4\,D_6} \left(\,(1-\epsilon)\,\alpha_1^2\, \bm{A}_1\,\bm{B}_3 - \alpha_1^2\,\alpha_2\,\bar{ \alpha }_1\,\bm{r}^2 \right)\\
\A\B_{14}^L=&{}\;
 N_c \;\frac{1}{D_1\,D_2\,D_5\,D_7}\,\bigg( \frac{1}{2\,
          \alpha_\ell}\,\bar{ \alpha }_1\,
          \left( \alpha_1\,\bar{ \alpha }_1 -\alpha_\ell - (1-\epsilon)\,2\,\alpha_\ell^2 \right) \bm{A}_1\,\bm{B}_4 \nonumber \\*
     &+ \alpha_1\,\bar{ \alpha }_1^2\,
        \bar{ \alpha }_2\,D_2 - 
       \frac{1}{2}\alpha_1\,\alpha_2\,\bar{ \alpha }_1^2\,
          D_7 + 
       \frac{1}{2\,\alpha_\ell}\,\alpha_1\,\alpha_2\,
          \bar{ \alpha }_1\,\bar{ \alpha }_2\, \bm{A}_1 \bm{r}- 
       \alpha_1\,\alpha_2\,\bar{ \alpha }_1\,\bar{ \alpha }_2\,
        \bm{r}^2 \nonumber \\*
      &- \frac{3}{2}\,\bar{ \alpha }_1\,\bar{ \alpha }_2\,
          \bm{A}_1 \left( \bm{k} + \bm{r} \right)\,\bigg) \\
\A\B_{15}^L=&{}\;
   - N_c \;\frac{1}{\alpha_\ell\,D_1\,D_2\,D_5\,D_7}\alpha_1\,\alpha_2\,\bar{ \alpha }_1^2\,
      \bar{ \alpha }_2\,\bm{r}^2  \\
\A\B_{22}^L=&{}\;
 \frac{1}{N_c}\;\frac{1}{D_2\,D_3\,D_4\,D_7}\left((1-\epsilon)\,\alpha_2\,\bar{ \alpha }_2\, \bm{A}_2\,\bm{B}_2 + 
     \alpha_2\,\bar{ \alpha }_2\,\left( \bm{k} + \bm{r} - \alpha_1\,\bm{r} \right) \,
      \left( \bm{k} + \bm{\ell} + \bar{ \alpha }_1 \,\bm{r} \right) \right) \\
\A\B_{23}^L=&{}\;
   - C_F \;\frac{2}{\bar{ \alpha
  }_1\,D_2\,D_4^2\,D_6}\,\,(1-\epsilon)\,\alpha_1^2\,\alpha_2\,\bm{A}_2 \bm{B}_3\\
\A\B_{24}^L=&{}\;
   - N_c\;\frac{1}{D_2\,D_4\,D_5\,D_7}\,\bigg( \frac{1}{2\,\alpha_\ell} \,\alpha_2\,
            \left( \alpha_1\,\bar{ \alpha }_1 -\alpha_\ell - 2\,(1-\epsilon)\,\alpha_\ell^2 \right) \, \bm{A}_2 \,\bm{B}_4 \nonumber\\*  
  & + \alpha_1\,\alpha_2\,\bar{ \alpha }_1\,\bar{ \alpha }_2\,
        D_2 - \frac{1}{2} \alpha_1\,\alpha_2^2\,
          \bar{ \alpha }_1\,D_7 + 
       \frac{1}{2\,\alpha_\ell} \,\alpha_1\,\alpha_2\,
          \bar{ \alpha }_1\,\bar{ \alpha }_2\, \bm{A}_2 \bm{r} + 
       \alpha_1\,\alpha_2\,\bar{ \alpha }_1\,\bar{ \alpha }_2\,
        \bm{r}^2 \nonumber \\* 
   &- \frac{3}{2} \,\alpha_2\,\bar{ \alpha }_2\,
          \bm{A}_2 \left( \bm{k} + \bm{r} \right)\,\bigg) \\
\A\B_{25}^L=&{}\;
 N_c\;\frac{1}{\alpha_\ell\,D_2\,D_4\,D_5\,D_7}\alpha_1\,\alpha_2^2\,\bar{ \alpha }_1\,\bar{ \alpha }_2\,\bm{r}^2\\
\A\B_{33}^L=&{}\;
 \frac{1}{N_c}\; \frac{1}{D_1\,D_3\,D_4\,D_6}\left((1-\epsilon)\,\alpha_1\,\bar{ \alpha }_1\, \bm{A}_3\,\bm{B}_3 + 
     \alpha_1\,\bar{ \alpha }_1\,\left( \bm{k} + \bm{r} - \alpha_1\,\bm{r} \right) \,
      \left( \bm{k} + \bm{\ell} + \bar{ \alpha }_1 \,\bm{r} \right) \right) \\
\A\B_{34}^L=&{}\;
   - N_c \,\frac{1} {D_1\,D_3\,D_5\,D_7}\,
       \bigg( \frac{1}{2\,\alpha_1\,\alpha_\ell}\bar{ \alpha }_1^2\,
            \left( \alpha_1\,\bar{ \alpha }_2 + 2\,(1-\epsilon)\,\alpha_\ell^2  
              \right) \,\bm{A}_3\,
            \bm{B}_4 \nonumber \\*
     &+ \frac{1}{2}\alpha_1\,\bar{ \alpha }_1^2\,\bar{ \alpha }_2\,D_7 + 
         \frac{1}{2\,\alpha_\ell}\alpha_1\,\bar{ \alpha }_1^2\,\bar{
           \alpha }_2\,\bm{A}_3 \bm{r} + 
         \alpha_1\,\bar{ \alpha }_1^2\,\bar{ \alpha }_2\,\bm{r}^2 \bigg)\\
\A\B_{35}^L=&{}\;
 N_c \;\frac{1}{\alpha_\ell\,D_1\,D_3\,D_5\,D_7}\alpha_1\,\bar{ \alpha }_1^2\,\bar{ \alpha }_2^2\,\bm{r}^2\\
\A\B_{44}^L=&{}\;
   - N_c \,\frac{1}{D_2\,D_5^2\,D_7} \,\bigg( \left(  2\,\alpha_1\,\bar{ \alpha }_1 + \alpha_\ell - 2\,(1-\epsilon)\,\alpha_\ell^2\right)  \bm{A}_4\,\bm{B}_4 \nonumber \\*
     &+ \frac{1}{2} \alpha_1\,\alpha_\ell\,\bar{ \alpha }_1\,
          \bar{ \alpha }_2\,D_2 + 
       \frac{1}{2} \alpha_1\,\alpha_2\,\alpha_\ell\,
          \bar{ \alpha }_1\,D_7 +  
    2\,\alpha_1\,\alpha_2\,\bar{ \alpha }_1\,\bar{ \alpha }_2\,
       \bm{r}^2 - \alpha_\ell^2\,\left( \bm{k} + \bm{\ell} \right)
   \,\left( \bm{k} + \bm{r} \right) 
       \bigg) \\
\A\B_{45}^L=&{}\;
  N_c\;\frac{1}{D_2\,D_5^2\,D_7}\alpha_1\,\alpha_2\,\bar{ \alpha
   }_1\,\bar{ \alpha }_2\,\bm{r}^2  \\
\A\B_{55}^L=&{}\;0 \,.
\label{eq:AB55}
\end{align}
Here we have used the following abbreviations:
\begin{align}
\bm{A}_1=&\,\alpha_2\, \bm{\ell} + \alpha_\ell \, (\bm{k}+\bm{r})\\
\bm{A}_2=&\,\bar{\alpha}_1\, \bm{\ell} + \alpha_\ell \, (\bm{k}+\bm{\ell})\\
\bm{A}_3=&\,\alpha_1\, \bm{\ell} - \alpha_\ell \, (\bm{k}+\bm{r})\\ 
\bm{A}_4=&\,\alpha_2\, (\bm{\ell} - \bm r) + \alpha_\ell \, (\bm{k}+\bm{r})\\
\bm{B}_1=&\,\bar{\alpha}_2\, \bm{\ell} - \alpha_\ell \,
 (\bm{k}+\bm{\ell})\\
\bm{B}_2=&\,\alpha_1\, \bm{\ell} - \alpha_\ell \, (\bm{k}+\bm{r})\\
\bm{B}_3=&\,\bar{\alpha}_1\, \bm{\ell} + \alpha_\ell \,
 (\bm{k}+\bm{\ell})\\
\bm{B}_4=&\,\bar{\alpha}_2\, (\bm{\ell} - \bm r) - \alpha_\ell \,
(\bm{k}+\bm{\ell})\,.
\label{eq:b4def}
\end{align}

We conclude our listing of the squared matrix elements with a few remarks.
First, similar to the longitudinal case, the sum of the squared matrix 
elements for 
the transverse photon is ultraviolet finite: individual terms diverge as 
$\bm{\ell}$ becomes large (for fixed $\bm{k}$), but in the sum these
divergences cancel. Secondly, it is easy to show that the factorization 
properties in configuration space which for the longitudinal photon 
have been discussed in section~V of 
\cite{BGK} also hold for the transverse photon: this supports the validity  
of the photon wave function picture beyond the LO approximation. 
Finally, the limit of small $\alpha_{\ell}$
agrees with the LO BFKL calculation; we shall return to this question 
further below.  

\section{Singular Limits of the Real Corrections}
\label{sec:singular-real}

\noindent
In order to formulate finite expressions for the $q\bar{q}g$
contribution to the photon impact factor we examine the divergent
limits of the squared matrix element $|\M_{q\bar qg}|^2$ for the
process $\gs + q \to q\bar qg+q$.  For the phase space of the $q\bar
qg$-final state we have from (\ref{eq:ps3}), (\ref{eq:psg}):
\begin{equation}
  \label{eq:qqgps}
  d\phi_{q\bar qg} \frac{sd\beta_r}{2\pi} = 
  \frac{d\alpha}{2 \alpha (1-\alpha-\alpha_\ell)} 
  \frac{d^{D-2}\bm k}{(2\pi)^{D-1}}     
  \frac{d\alpha_\ell}{2 \alpha_\ell} 
  \frac{d^{D-2}\bm \ell}{(2\pi)^{D-1}} 
\end{equation}
In the following we suppress the superscript $(D-2)$ in the
integration measure of the transverse momenta. The contribution of the
$q\bar{q}g$ state to the impact factor reads:
\begin{equation}
\label{eq:impqqg}
\int \frac{d\alpha}{2 \alpha (1-\alpha-\alpha_\ell)} 
  \int \frac{d \bm k}{(2\pi)^{D-1}}     
  \int \frac{d\alpha_\ell}{2 \alpha_\ell} 
  \int \frac{d \bm \ell}{(2\pi)^{D-1}}\;\;\;  |\Gamma_{\gs\to q\bar qg}^{(0)}|^2, 
\end{equation}
where the square of the vertex function, $|\Gamma_{\gs\to q\bar qg}^{(0)}|^2$,
follows from (\ref{eq:m2l}) and (\ref{eq:m2t}).
   
We begin by examining the limit in which the gluon is either collinear
to the quark or to the antiquark of the $q\bar qg$-system.  In the latter
case, the gluon momentum $\ell$ is collinear with the momentum $q-k-\ell$
of the antiquark, resulting in $D_4 = 0$ or 
\begin{equation}
  \bm \ell' = \bm\ell + \frac{\alpha_\ell}{1-\alpha} \bm k = 0\; .
\end{equation}
We approximate our results for $|\Gamma_{\gs\to q\bar qg}^{(0), a}|^2$
in the region around the singularity, i.e.\ we calculate the residue
of the pole at $D_4 = 0$.  The collinear limit of QCD matrix elements is
well known and can be taken from the literature \cite{CSdiv}.  
As an additional check, we have calculated the
divergent limits directly from our matrix elements.
For both polarizations of the photon, the $g\parallel\bar q$
collinear limit of (\ref{eq:impqqg}) can be written as
\begin{align}
  &\left.|\Gamma_{\gs\to q\bar qg}^{(0)}|^2 
    d\phi_{q\bar qg} \frac{sd\beta_r}{2\pi}\right|^{{\mathrm coll,}\bar q}
  = \,|\Gamma_{\gs\to q\bar q}^{(0)}|^2 
  \frac{d\alpha}{2 \alpha (1-\alpha)} 
  \frac{d\bm k}{(2\pi)^{3-2\epsilon}}   \nonumber \\
  &\qquad\quad\times C_F 
  P_{gq}\left(\frac{\alpha_\ell}{1-\alpha}, \epsilon \right)
  \frac{d\alpha_\ell}{(1-\alpha)} 
  \frac{d\bm\ell}{(2\pi)^{3-2\epsilon}{\bm{\ell'}}^2}\\
  \label{eq:colla}
  &\qquad= \,\I_2 (\alpha, \bm k) d\alpha\,d\bm k \,
  C_F P_{gq}\left(\frac{\alpha_\ell}{1-\alpha}, \epsilon \right)
  \frac{d\alpha_\ell}{(1-\alpha)}
  \frac{d\bm\ell}{(2\pi)^{3-2\epsilon}{\bm{\ell'}}^2}
\end{align}
where $\I_2 (\alpha, \bm k)\equiv\I_2 (\alpha,\bm k;\bm r,Q)$ was defined in
(\ref{eq:bornif}), (\ref{eq:bornifl}), (\ref{eq:bornift}), and 
$C_F P_{gq}(z,\epsilon)$ denotes the $q\to g$
Altarelli-Parisi splitting function in $D=4-2\epsilon$ dimensions:
\begin{equation}
  \label{eq:apsplitting}
  P_{gq}(z, \epsilon) = \frac{1 + (1-z)^2}{z} - \epsilon z\,.
\end{equation}
The color coefficient $C_F$ indicates that, for the examination of 
singular limits it will be convenient to use the color basis $C_F$ and 
$C_A$ rather than the color factors $N_c$, $C_F$, and $1/N_c$ which 
appear in the matrix elements of the previous section. In this basis, the
$C_A$ part is free from collinear singularities.

In the case where the gluon is collinear to the quark, the infrared limit is
obtained from (\ref{eq:colla}) by substituting%
\footnote{This is nothing but the $q\leftrightarrow\bar q$ variables exchange at
  fixed gluon momentum.}
$\alpha$ by $\bar\alpha = 1 - \alpha - \alpha_\ell$ and $\bm k$ by 
$\bar{\bm k} = -\bm k-\bm r -\bm\ell$. The ``soft-collinear'' limit of
(\ref{eq:colla})  that is needed later is
obtained by retaining from the splitting function only the pole $2/z$:
\begin{equation}
  \label{eq:s+c}
  \left.|\Gamma_{\gs\to q\bar qg}^{(0)}|^2 
   d\phi_{q\bar qg} \frac{sd\beta_r}{2\pi}\right|^{{\mathrm soft,coll,}\bar q}
  = \,\I_2 (\alpha, \bm k) d\alpha\,d\bm k \,
  2C_F\,\frac{d\alpha_\ell}{\alpha_\ell} 
  \frac{d\bm \ell}{(2\pi)^{3-2\epsilon}{\bm{\ell'}}^2}\,.
\end{equation}

Next, we examine the `soft limit' where the gluon momentum
$\alpha_\ell\to0$ and $\bm\ell\to0$. Formally we substitute $\ell \to
\rho\ell$ or $\bm\ell\to\rho\bm\ell,\alpha_\ell\to\rho\alpha_\ell$,
expand around $\rho=0$ and keep only the leading term.  As in the
collinear case, the result is proportional to the Born approximation,
for both polarizations of the photon. We separate the result according
to different colour structures:
\begin{equation}
  \label{eq:softqqg}
  \left.|\Gamma_{\gs\to q\bar qg}^{(0)}|^2 \right|^{\mathrm{soft}}
  =   C_A\, \left.|\Gamma_{\gs\to q\bar qg}^{(0)}|^2
  \right|^{\mathrm{soft}}_{C_A}
  + C_F \left.|\Gamma_{\gs\to q\bar qg}^{(0)}|^2
  \right|^{\mathrm{soft}}_{C_F} 
\end{equation}
with 
\begin{equation}
  \label{eq:softqqgA}
  \left.|\Gamma_{\gs\to q\bar qg}^{(0)}|^2 \right|^{\mathrm{soft}}_{C_A}
  = \,|\Gamma_{\gs\to q\bar q}^{(0)}|^2 
  \frac{4\left(\bm\ell - \alpha_\ell \bm a \right)
    \cdot \left(\bm\ell + \alpha_\ell \bm b \right)}
  {\left(\bm\ell - \alpha_\ell \bm a\right)^2 
    \left(\bm\ell + \alpha_\ell \bm b \right)^2}\;  
\end{equation}
and 
\begin{equation}
  \label{eq:softqqgF}
  \left.|\Gamma_{\gs\to q\bar qg}^{(0)}|^2 \right|^{\mathrm{soft}}_{C_F}
  = \,|\Gamma_{\gs\to q\bar q}^{(0)}|^2 
  \frac{4\,\alpha_\ell^2(\bm a + \bm b)^2}
  {\left(\bm\ell - \alpha_\ell \bm a\right)^2 
    \left(\bm\ell + \alpha_\ell \bm b \right)^2}\,,  
\end{equation}
where, for convenience, we have defined 
\begin{equation}
  \bm a = \frac{\bm k + \bm r}{\alpha}, \qquad \bm b = \frac{\bm k}{1-\alpha}.
\end{equation}
In the soft limit, the phase space measure (\ref{eq:qqgps}) factorizes in the
product of the $q\bar q$ measure (\ref{eq:qqps}) and differentials of the 
gluon variables
\begin{equation}\label{eq:softps}
 d\phi_{q\bar qg} \frac{sd\beta_r}{2\pi} \to \left(
 d\phi_{q\bar q} \frac{sd\beta_r}{2\pi} \right) \left( 
 \frac{d\alpha_\ell}{2 \alpha_\ell}\frac{d\bm \ell}{(2\pi)^{D-1}}\right) \,.
\end{equation}
Our result for the soft limit agrees with \cite{FM}. We also note that, 
in the limit $\bm \ell' = \bm\ell
+\alpha_\ell \bm b \to 0$,  we reproduce  from
(\ref{eq:softqqgF}) and (\ref{eq:softps}) the ``soft-collinear'' limit 
(\ref{eq:s+c}).

Finally, we have to consider the limit in which the gluon is emitted
in the central rapidity region. Formally, this corresponds to the
singular limit $\alpha_{\ell} \to 0$: the matrix elements listed in
the previous section are finite for $\alpha_{\ell} \to 0$, but the
phase space element (\ref{eq:psg}) introduces a logarithmic
singularity.  Consequently, when doing the subsequent $\alpha_{\ell}$
integration this divergence $\sim 1/\alpha_{\ell}$ will produce the
leading-$\ln s$ logarithm.  As discussed in
section~\ref{sec:definition}, this part of the gluon phase space
belongs to final states with two large rapidity gaps and has to be
subtracted from our result. This will be the topic of
section~\ref{sec:NLOIF}. For the moment we only state that in the
limit $\alpha_{\ell} \to 0$, for both polarizations of the photon, the
sums of our matrix elements split into the product of the Born
approximation and the BFKL kernel:
\begin{equation}
  \label{eq:LLlimit}
  \left.|\Gamma_{\gs\to q\bar qg}^{(0)}|^2\right|_{\mathrm{LL}}
  = \,\left.|\Gamma_{\gs\to q\bar q}^{(0)}|^2
 \right|_{\mscr{\begin{array}[b]{l}
       \bm r\to\bm r-\bm\ell \\
       \bm k\to\bm k+\bm\ell
     \end{array}}}
  \,\frac{1}{\left(\bm\ell - \bm r\right)^4}
  \frac{4C_A \bm r^2 \left(\bm\ell - \bm r\right)^2}{\bm\ell^2}\,.
\end{equation}
Here the last factor is the BFKL kernel (without the gluon trajectory 
function)
\begin{equation}\label{eq:kernel}
 \K(\bm r-\bm\ell,\bm r) = \frac{C_A}{\pi}
 \frac{\bm r^2 \left(\bm\ell - \bm r\right)^2}{\bm\ell^2}, 
\end{equation}
and $1/\left(\bm\ell - \bm r\right)^4$ are the propagators of the 
(reggeized) gluons.  Note that in the Born level vertex functions the
transverse momenta are shifted from $\bm r$ to 
$\bm r-\bm\ell$ and from $\bm k$ to $\bm k + \bm \ell$. The color factor 
$C_A$ signals that the $C_F$ part of our matrix elements does not contribute 
to our subtraction of the central region.  In the limit where the
centrally emitted gluon is soft ($\bm\ell\ll \bm r$), we obtain
\begin{equation}
  \label{eq:LLsoftlimit}
  \left.|\Gamma_{\gs\to q\bar qg}^{(0)}|^2\right|_{\mathrm{LL}}^{\mathrm soft}
  = \,|\Gamma_{\gs\to q\bar q}^{(0)}(\alpha, \bm k)|^2 
  \frac{4C_A}{\bm\ell^2}\,.
\end{equation}
The same result is obtained from the soft limit (\ref{eq:softqqgA}),
if we go into the central region limit $\alpha_\ell \ll 1$.

\section{Subtraction of the Central Region}
\label{sec:NLOIF}

\noindent
After having calculated the various divergent limits of our matrix elements, 
we now turn to the subtraction of the central region. 
As we have discussed in section~\ref{sec:definition}, contributions to 
the total cross section are divided according to the number of large 
rapidity gaps. What we have presented in the last two sections are 
the results for the process $\gamma^* +q \to (q\bar{q}g)+q$, imposing the  
constraint of a large rapidity gap between the $(q\bar{q}g)$ system 
and the quark. These results therefore contain, as a special case, 
still the configuration where the $q\bar{q}g$ system contains a second 
rapidity gap 
between the $q\bar{q}$ pair and the gluon: this piece has to be removed     
since it counts as a configuration with two gaps and, in order 
$\alpha_s^3$ of the total cross section, it entirely belongs to the 
leading-$\ln s$ approximation (\ref{eq:svil}).

In order to make the separation more explicit, we first divide the phase space 
of the produced gluon into 'upper' and 'lower' halves. Introducing 
\begin{equation}
\alpha_{\ell}^{cut}=\frac{|\bm \ell|}{\sqrt{s}},   
\end{equation}
the 'upper' region $\alpha_{\ell}^{cut}< \alpha_{\ell} < 1-\alpha$ consists of 
the 'upper' half of the central region plus the fragmentation region of the 
incoming photon, whereas the 'lower' region $\alpha_{\ell} <  
\alpha_{\ell}^{cut}$ contains the lower half of the central region plus the 
fragmentation region of the quark. 
In the upper region we write:
\begin{align}
\label{eq:intGO}
&\int |\Gamma_{\gs\to q\bar qg}^{(0)}|^2 d\phi_{q\bar q g} \frac{sd\beta_r}{2\pi}
  \Theta(\alpha_\ell - \cut) \nonumber \\
&= \int d\bm k\,d\bm\ell\int_0^1 d\alpha \int_0^{1-\alpha} d\alpha_\ell\,
 \I_3(\alpha_\ell, \bm \ell; \alpha, \bm k; \bm r, Q)\Theta(\alpha_\ell-\cut)
\end{align}
Here $\I_3$ denotes the full integrand involving the sum of the matrix 
elements  of the three-particle final state.

The division into central region and fragmentation region is closely
related to the choice of the energy scale $s_0$. In (\ref{eq:svil}) we
have introduced the energy scale $s_0$: since a change in $s_0$ can be
absorbed into the NLO impact factors $\Phi_A^{(1)}$ and
$\Phi_B^{(1)}$, one might expect that there exists some freedom in
choosing this scale. We adopt the simplest choice, namely a constant
value which is independent of kinematic variables inside the impact
factors. Examples for $s_0$ include the (negative) virtuality of the
$t$-channel gluon, $\bm r^2$, or the (negative) mass of the photon,
$Q^2$.

In order to define the $q\bar{q}g$ contribution to the photon impact
factor, $\Phi_{\gamma^*}^{(1,\mathrm{real})}$, we have to subtract,
starting from (\ref{eq:intGO}), the leading-$\ln s$ piece with the
energy scale $s_0$. This can be achieved by considering the central
region limit~(\ref{eq:LLlimit}) of the matrix element which implies
\begin{equation}\label{eq:I3cr}
 \left.\I_3(\alpha_\ell, \bm \ell; \alpha, \bm k; \bm r, Q)\right|_{\mathrm LL}
 =\I_2(\alpha, \bm k+\bm\ell; \bm r-\bm l,Q)
 \frac1{(2\pi)^{2-2\epsilon}(\bm\ell-\bm r)^4}
 \K(\bm r-\bm\ell,\bm r)\frac1{\alpha_\ell}\,.
\end{equation}
We recognize the integrand $\I_2$ occurring in the LO impact
factor~(\ref{eq:bornif}) with the value of the reggeon momentum being
shifted to $\bm r \to \bm r - \bm \ell$ (since the momentum $\bm \ell$
is carried by the outgoing gluon), the propagators associated to the
$t$-channel (reggeized) gluons, the measure factor
$1/(2\pi)^{2-2\epsilon}$ required by the
definition~(\ref{eq:factorized}), the real part of the BFKL kernel,
and, finally, a factor $1/\alpha_\ell$.  It is this factor
$1/\alpha_{\ell}$ which, when the integration over $\alpha_\ell$ is
done, provides the logarithm of the energy.  Our aim of finding the LL
piece with energy scale $s_0$ leads to the following $\alpha_{\ell}$
integral:
\begin{equation}
 \int_0^{\amax} \frac{d\alpha_\ell}{\alpha_\ell}\Theta(\alpha_\ell - \cut)
 = \frac12 \log\frac{(\amax)^2 s}{\bm\ell^2}\,.
\end{equation}
with $\amax\equiv|\bm\ell|/\sqrt{s_0}$. Therefore, we define as genuine real
emission contribution to the photon impact factor the expression
\begin{align}\nonumber
 \Phi_{\gs}^{(1,{\mathrm real})}
&=\int d\bm k\,d\bm\ell\int_0^1 d\alpha \int_{\cut}^\infty d\alpha_\ell\,\Big[
 \I_3(\alpha_\ell, \bm \ell; \alpha, \bm k; \bm r, Q)\Theta(1-\alpha-\alpha_\ell) \\
 &\quad-\I_2(\alpha, \bm k+\bm\ell; \bm r-\bm l,Q)
 \frac1{(2\pi)^{2-2\epsilon}(\bm\ell-\bm r)^4}
 \K(\bm r-\bm\ell,\bm r)\frac1{\alpha_\ell}
 \Theta(\frac{|\bm\ell|}{\sqrt{s_0}}-\alpha_\ell)\Big]\,. \label{eq:defIF}
\end{align}

In this definition of our impact factor, the following remark is in
place.  With our choice of a constant energy scale, which has led to
the subtraction, we are adding (in the LL term) and subtracting (in
$\Phi_{\gs}^{(1,{\mathrm real})}$) a region of the phase space which
is kinematically forbidden. In fact, the LL subtraction (last line of
Eq.~(\ref{eq:defIF})) includes the region of phase space where
$\alpha_\ell>1-\alpha$. It is important to make sure that this
``forbidden'' region
$|\bm\ell|>(1-\alpha)\sqrt{s_0}\,,\;1-\alpha<\alpha_\ell<|\bm\ell|/\sqrt{s_0}$
contributes a finite term and does not introduce new divergences.
The finiteness results from the ``good'' UV behaviour of the LO impact
factor, namely $\Phi_{\gs}^{(0)}(\bm r-\bm\ell)\to\text{const}$ for
$\bm\ell\to\infty$ (up to logarithms). After multiplication with the
propagator $\sim1/\bm\ell^4$ and with the factor
$\log[|\bm\ell|/(1-\alpha)\sqrt{s_0}]$ (which is due to the
$\alpha_\ell$ integral), the resulting $\bm\ell$ integral is finite.
This situation is quite different from the case of partonic impact
factors \cite{CiCo-if,Fadin-gif}, where $\Phi_{q,g}^{(0)}(\bm r-\bm\ell)\sim
\bm\ell^2$, and the $\bm\ell$ integral, after including the gluon
propagators and the logarithmic factor, leads to a double pole in
$1/\epsilon$.

\section{Finite Combinations}
\label{sec:finite}

\noindent
In the previous section we have suggested a suitable form of the LL
subtraction and thus specified the real emission contribution to the
impact factor. What remains are the infrared divergences which have to cancel 
once we combine real and virtual corrections. In this final section of our 
paper we show explicitly the cancellation of all the IR singularities, and we 
present expressions consisting of well defined finite terms.

Given that all the singular terms of the virtual corrections are proportional to
the LO impact factor~(\ref{eq:bornif}), as one can see from
Eq.~(\ref{eq:T1div}), it is convenient to study the singular regions of the
phase space integrals of the real corrections by integrating first only over the
gluon variables. In fact, the gluon integration generates singular terms which
are in turn proportional to the LO impact factor. They will be combined with the
singular virtual corrections after a proper identification of the remaining
integration variables.

It is also convenient to separate the different colour structure contributions
into a $C_A$ term and a $C_F$ term
\begin{equation}
 \I_3\equiv C_A \I_3^{C_A} + C_F \I_3^{C_F}\,,
\end{equation}
because the LL subtraction is contained only in the former term, whilst the
collinear singularities are present only in the latter. Soft singularities are
found in both terms.

\subsection{$C_A$ term}\label{ss:ca}

\noindent
Let us consider the $C_A$ contribution to the real emission impact factor
corrections, as specified in Eq.~(\ref{eq:defIF}). We want to compute first
the $\bm\ell$ and $\alpha_\ell$ integrals at fixed $\bm k$ and $\alpha$:
\begin{align}\nonumber
 \left.\frac{d\Phi_{\gs}^{(1)}}{d\alpha\,d\bm k}\right|_{C_A}=
\int d\bm\ell\int_{\cut}^\infty d\alpha_\ell\,&[
 \I_3^{C_A}(\alpha_\ell, \bm \ell; \alpha, \bm k; \bm r, Q)
 \Theta(1-\alpha-\alpha_\ell) \\
&-\I_2(\alpha, \bm k+\bm\ell;\bm r-\bm l,Q)\frac2{(2\pi)^{3-2\epsilon}}
 \frac{\bm r^2}{(\bm r-\bm\ell)^2 \bm\ell^2}\frac1{\alpha_\ell}
 \Theta(\frac{|\bm\ell|}{\sqrt{s_0}}-\alpha_\ell)]\,. \label{eq:CA}
\end{align}

This integral is divergent in the soft region
$\alpha_\ell\to0,\;\bm\ell\to0$ when $\epsilon=0$.  We evaluate it in
dimensional regularization by means of the subtraction method: we
subtract from the integrand its soft approximation, in such a way that
the resulting integral will be $\epsilon$-finite. Subsequently we re-add 
what we have subtracted, and we integrate this term analytically.
Of course, the soft limit is not uniquely defined outside the soft
region. We choose the soft approximation according to
Eqs.~(\ref{eq:softqqgA}, \ref{eq:LLsoftlimit}, \ref{eq:softps}); we
extend the soft subtraction up to the kinematic limit of the gluon
longitudinal momentum, i.e., $\alpha_\ell<1-\alpha$:
\begin{align}\nonumber
 \left.\frac{d\Phi_{\gs}^{(1)}}{d\alpha\,d\bm k}\right|_{C_A}^{\mathrm soft}&=
 \I_2(\alpha, \bm k; \bm r,Q)\frac2{(2\pi)^{3-2\epsilon}}\\ \label{eq:CAsoft}
&\quad\times \int d\bm\ell\int_{\cut}^\infty \frac{d\alpha_\ell}{\alpha_\ell}
\left[\frac{\left(\bm\ell - \alpha_\ell \bm a \right)
    \cdot \left(\bm\ell + \alpha_\ell \bm b \right)}
  {\left(\bm\ell - \alpha_\ell \bm a\right)^2 
    \left(\bm\ell + \alpha_\ell \bm b \right)^2}
  -\frac1{\bm\ell^2}\Theta(\frac{|\bm\ell|}{\sqrt{s_0}}-\alpha_\ell)\right]
  \Theta(1-\alpha-\alpha_\ell)\,.
\end{align}
Accordingly, we define the finite $C_A$ part of the real corrections to the impact factor as
\begin{align}\nonumber
 &\left.\Phi_{\gs}^{(1,{\mathrm real})}\right|_{C_A}^{\mathrm finite}\equiv
 \int d\bm k\, d\bm\ell\int_0^1 d\alpha\int_0^\infty d\alpha_\ell
 \Biggl\{[\I_3^{C_A}(\alpha_\ell, \bm \ell; \alpha, \bm k; \bm r, Q)
 \Theta(1-\alpha-\alpha_\ell) \\ \label{eq:CAfinite}
&-\I_2(\alpha, \bm k+\bm\ell; \bm r-\bm l,Q)\frac2{(2\pi)^{3-2\epsilon}}
 \frac{\bm r^2}{(\bm r-\bm\ell)^2 \bm\ell^2}\frac1{\alpha_\ell}
 \Theta(\frac{|\bm\ell|}{\sqrt{s_0}}-\alpha_\ell)] \\ \nonumber
 &-\I_2(\alpha, \bm k; \bm r,Q)\frac2{(2\pi)^{3-2\epsilon}}\frac1{\alpha_\ell}
\left[\frac{\left(\bm\ell - \alpha_\ell \bm a \right)
    \cdot \left(\bm\ell + \alpha_\ell \bm b \right)}
  {\left(\bm\ell - \alpha_\ell \bm a\right)^2 
    \left(\bm\ell + \alpha_\ell \bm b \right)^2}
  -\frac1{\bm\ell^2}\Theta(\frac{|\bm\ell|}{\sqrt{s_0}}-\alpha_\ell)\right]
 \Theta(1-\alpha-\alpha_\ell)\Biggr\}\,.
\end{align}
Note that we have shifted to zero the lower limit of integration of
$\alpha_\ell$ because, thanks to the LL subtraction, there is no
contribution to the integral coming from the central region.
Note also that the singularities of the integrand due to the
denominators $\bm\ell-\alpha_\ell\bm a$ and $\bm\ell+\alpha_\ell\bm b$
are integrable, because of the scalar product in the
numerator. This reflects the fact that the $C_A$ part is free of
collinear singularities.

The expression written in Eq.~(\ref{eq:CAsoft}) can be computed
analytically.  Since the lower limit of integration in $\alpha_\ell$
depends on the energy $s$, one gets an $s$ dependent result. However,
because of the LL-soft subtraction to the soft term, the central
region contribution is suppressed, and we expect a finite expression
in the $s\to\infty$ limit.  We can compute the integrals of the two
terms in Eq.~(\ref{eq:CAsoft}) in this limit ($\cut\to 0$) separately
because dimensional regularization promotes the divergent $\log s$
contributions into UV $\epsilon$-poles, which will cancel in the difference.
We have:
\begin{equation}\label{integral1}
 J_1\equiv\int d\bm\ell\int_0^{1-\alpha} \frac{d\alpha_\ell}{\alpha_\ell}
 \frac{\left(\bm\ell - \alpha_\ell \bm a \right)
    \cdot \left(\bm\ell + \alpha_\ell \bm b \right)}
  {\left(\bm\ell - \alpha_\ell \bm a\right)^2 
    \left(\bm\ell + \alpha_\ell \bm b \right)^2}
 = -c_\Gamma \pi^{1-\epsilon}\frac{(1-\alpha)^{-2\epsilon}}{2\epsilon^2}
 \left((\bm a+\bm b)^2\right)^{-\epsilon}
\end{equation}
and
\begin{equation}\label{integral2}
 J_2\equiv\int d\bm\ell\int_0^{1-\alpha}
 \frac{d\alpha_\ell}{\alpha_\ell}\frac1{\bm\ell^2}  
 \Theta(\frac{|\bm\ell|}{\sqrt{s_0}}-\alpha_\ell)
 = -\frac{\pi^{1-\epsilon}}{\Gamma(1-\epsilon)}
 \frac{(1-\alpha)^{-2\epsilon}}{2\epsilon^2}s_0^{-\epsilon}\,.
\end{equation}
These two integrals have to be multiplied by $\I_2$ and integrated
over $\alpha$ and $\bm k$. After Eq.~(\ref{eq:bornif}) we have noticed
that $\I_2$ is symmetric under the change of variables
$\alpha\leftrightarrow1-\alpha$, $\bm k\leftrightarrow-\bm\ell-\bm k$,
corresponding to the $q\leftrightarrow\bar q$ exchange; the term $(\bm
a+\bm b)^2=M^2/\alpha(1-\alpha)$ is also invariant under this
exchange.  Therefore, on the right-hand-side of Eqs.~(\ref{integral1})
and (\ref{integral2}), we can replace
$(1-\alpha)^{-2\epsilon}\to\frac12\alpha^{-2\epsilon}
+\frac12(1-\alpha)^{-2\epsilon}$, which casts the expressions into
explicitly symmetric forms w.r.t.\ $\alpha\leftrightarrow1-\alpha$.
This manipulation will be useful when combining real and virtual
corrections.

In conclusion, the divergent $C_A$ part of the real corrections to the impact
factor is
\begin{align}\nonumber
 \left.\Phi_{\gs}^{(1,{\mathrm real})}\right|_{C_A}^{\mathrm divergent}
 &\equiv\int d\bm k\int_0^1d\alpha\,
 \I_2(\alpha, \bm k; \bm r,Q)\frac2{(2\pi)^{3-2\epsilon}}(J_1-J_2)
 \\ \nonumber
&=\int d\bm k\int_0^1d\alpha\,\I_2(\alpha, \bm k; \bm r,Q)
 \frac{c_\Gamma}{(4\pi)^{2-\epsilon}}\biggl\{
 \frac{2}{\epsilon}\left[\log M^2-\log s_0-\log\alpha(1-\alpha)\right]
 \\ \label{eq:CAdiv}
 &\quad+\left[\left(\log s_0+\log\alpha(1-\alpha)\right)^2
 -\log^2 M^2\right]+\O(\epsilon)\biggr\}\,.
\end{align}
For consistency, the variables $\alpha$ and $\bm k$ of the above
equation have to be identified with the analogous variables of
Eq.~(\ref{eq:Phi1virtSing}).

\subsection{$C_F$ term}\label{ss:cf}

\noindent
Let us now consider the $C_F$ contribution to the real emission impact factor
corrections.  The first thing to note is that, since the BFKL kernel is
proportional to $C_A$, the LL subtraction of Eq.~(\ref{eq:defIF}) does not
affect the $C_F$ part. Therefore, the whole $C_F$ part of $\I_3$ contributes to
the impact factor. In addition, we can shift the lower limit $\cut$ for the
$\alpha_\ell$ integral to zero, just because the central region does not
contribute in this case.

The $C_F$ term contains both soft and collinear divergences. We separate the
singular contributions by means of the subtraction method. In what follows, the
last two parameters of $\I_2$ are always $\bm r,Q$. Therefore we simplify the
notation by indicating $\I_2(\alpha,\bm k)\equiv\I_2(\alpha,\bm k;\bm r,Q)$.

We start by identifying the soft singular term. This is done by subtracting 
from $\I_3^{C_F}$ its soft limit and then re-adding it again. We remark 
again that, beyond the soft region, the limit of integration is not uniquely 
defined. In
practice, by using the expression~(\ref{eq:softqqgF}), we adopt an
asymmetric function with respect to the $q\leftrightarrow\bar q$
exchange at fixed non-vanishing gluon momentum:
\begin{align}\nonumber
 \left.\frac{d\Phi_{\gs}^{(1)}}{d\alpha\,d\bm k}\right|_{C_F}^{\mathrm soft}&=
 \I_2(\alpha, \bm k)\frac{2(\bm a+\bm b)^2}{(2\pi)^{3-2\epsilon}}
 \int d\bm\ell\int_0^{1-\alpha}d\alpha_\ell\,
 \frac{\alpha_\ell}
  {\left(\bm\ell - \alpha_\ell \bm a\right)^2 
    \left(\bm\ell + \alpha_\ell \bm b \right)^2} \\ \label{eq:CFsoft}
&= \I_2(\alpha, \bm k)
 \frac{c_\Gamma}{(4\pi)^{2-\epsilon}}\frac{2}{\epsilon^2}
 \left[\alpha^{-2\epsilon}+(1-\alpha)^{-2\epsilon}\right]
 \left[\frac{M^2}{\alpha(1-\alpha)}\right]^{-\epsilon}\,,
\end{align}
where in the last equality a symmetrization in $\alpha$, as indicated after
Eq.~(\ref{integral2}), has been performed.

Both the full $C_F$ term $\propto\I_3^{C_F}$ and its soft
limit~(\ref{eq:CFsoft}) have a divergent behaviour when the gluon is either
collinear to the outgoing quark or antiquark.  Also for the evaluation of the
collinear divergences we adopt the subtraction method.  For the case of the 
gluon being collinear to the antiquark ($g\parallel\bar q$), we derive the 
limiting expression of the original integrand (i.e., without soft subtraction yet) from
Eq.~(\ref{eq:colla}), which yields ($\bm\ell'=\bm\ell+\alpha_\ell\bm b$)
\begin{align}\nonumber
 \left.\frac{d\Phi_{\gs}^{(1)}}{d\alpha\,d\bm k}\right|_{C_F}^{{\mathrm coll,}\bar q}&=
 \I_2(\alpha, \bm k)\int\frac{d\bm \ell}{(2\pi)^{3-2\epsilon}} 
 \int_0^{1-\alpha}\frac{d\alpha_\ell}{(1-\alpha)} 
 P_{gq}\left(\frac{\alpha_\ell}{1-\alpha}, \epsilon \right)
    \frac{\Theta(\alpha_\ell\Lambda - |\bm{\ell'}|)}{{\bm{\ell'}}^2}\nonumber \\
 &=\I_2(\alpha, \bm k)
 \,\frac{\alpha^{-2\epsilon}+(1-\alpha)^{-2\epsilon}}%
 {(4\pi)^{2-\epsilon}\Gamma(1-\epsilon)}\,
 \frac{\Lambda^{-2\epsilon}}{\epsilon}
 \left[\frac1{\epsilon}+\frac2{1-2\epsilon}-\frac12\right]
 \,,\label{eq:CFcollqbar}
\end{align}
where we have decided to perform the collinear subtraction only in a cone ---
specified by the $\Theta$ function --- whose axis coincides with the gluon
momentum being parallel to the antiquark momentum, and the vertex sits on the soft
point. The ``opening angle'' of the cone is parametrized by a cutoff $\Lambda$:
the bigger $\Lambda$, the wider the domain of the collinear subtraction.

In order to end up with a finite integral, we have to subtract also the
collinear singularities of the soft subtraction term~(\ref{eq:CFsoft}). For the
$g\parallel\bar q$ we have (see Eq.~(\ref{eq:s+c}):
\begin{align}\nonumber
 \left.\frac{d\Phi_{\gs}^{(1)}}{d\alpha\,d\bm k}
 \right|_{C_F}^{{\mathrm soft, coll,}\bar q}&=
 \I_2(\alpha, \bm k)\frac{2}{(2\pi)^{3-2\epsilon}}
 \int_0^{1-\alpha}\frac{d\alpha_\ell}{\alpha_\ell} \int d\bm{\ell}\,
  \frac{\Theta(\alpha_\ell\Lambda - |\bm{\ell'}|)}{{\bm{\ell'}}^2}
  \,.\\ \label{eq:CFsoftcollqbar}
&=\I_2(\alpha, \bm k)\frac1{(4\pi)^{2-\epsilon}\Gamma(1-\epsilon)}
 \left[\alpha^{-2\epsilon}+(1-\alpha)^{-2\epsilon}\right]
 \frac{\Lambda^{-2\epsilon}}{\epsilon^2}\,.
\end{align}
The soft-collinear subtraction uses the same cone as defined for the collinear
one. It is important that Eq.~(\ref{eq:CFsoftcollqbar}) can be obtained both as
a collinear limit of Eq.~(\ref{eq:CFsoft}) and as a soft limit of
Eq.~(\ref{eq:CFcollqbar}). This guarantees not only that the collinear
singularities introduced by the soft subtraction are subtracted, but also that
the soft singularities introduced by the collinear subtraction are properly
removed.

Note that the collinear term~(\ref{eq:CFcollqbar}) is proportional to
$\I_2(\alpha, \bm k)$.  This is a consequence of the convenient choice
of variable employed to parametrize the outgoing particles. In fact,
when the gluon is collinear to the antiquark, the $g\bar q$ system has
to be treated as a single entity carrying the sum of quantum numbers
and momenta of the constituting particles. In this case, the $g\bar q$
longitudinal momentum fraction and transverse momentum are $1-\alpha$
and $-\bm k$ respectively. The integration over the gluon variables
can therefore be independently performed leaving the LO integrand
$\I_2$ factored out.

The treatment of the $g\parallel q$ collinear singularity can be
performed in the same way as for the $g\parallel\bar q$ collinear
case. There are only two small differences: first of all, the
collinear limit of the integrand $\I_3$ is proportional to
$\I_2(\alpha+\alpha_\ell,\bm k+\bm\ell)$, which involves explicitly
the gluon variables $\alpha_\ell$ and $\bm\ell$. This is due to our
set of variables that does not describe the $qg$ system and the single
gluon independently. A more suitable set of variable in this case is
simply obtained by changing the $q\leftrightarrow\bar q$ labels:
\begin{subequations}\label{change}
\begin{align}
 \bar\alpha&\equiv1-\alpha-\alpha_\ell\\
 \bar{\bm k}&\equiv -\bm r-\bm k-\bm\ell\\
 \ellq&\equiv \bm\ell-\alpha_\ell\bm a\,.
\end{align}
\end{subequations}
From Eq.~(\ref{eq:CFcollqbar}) one gets
\begin{align}\nonumber
 \left.\frac{d\Phi_{\gs}^{(1)}}{d\bar\alpha\,d\bar{\bm k}}
 \right|_{C_F}^{{\mathrm coll,}q}
&=\I_2(\bar\alpha, \bar{\bm k})
 \int\frac{d\bm\ell}{(2\pi)^{3-2\epsilon}} 
 \int_0^{1-\bar\alpha}\frac{d\alpha_\ell}{(1-\bar\alpha)} 
 P_{gq}\left(\frac{\alpha_\ell}{1-\bar\alpha}, \epsilon \right)
 \\ \nonumber
&\quad\times\left(\frac{1-\bar\alpha}{1-\bar\alpha-\alpha_\ell}\right)^2
    \frac{\Theta(\alpha_\ell\Lambda 
 -\frac{1-\bar\alpha-\alpha_\ell}{1-\bar\alpha}|\ellq|)}{{\ellq}^2}\\
&=\I_2(1-\bar\alpha,-\bm r-\bar{\bm k})
 \,\frac{\bar\alpha^{-2\epsilon}+(1-\bar\alpha)^{-2\epsilon}}
 {(4\pi)^{2-\epsilon}\Gamma(1-\epsilon)}\,
 \frac{\Lambda^{-2\epsilon}}{\epsilon}
 \left[\frac1{\epsilon}+\frac2{1-2\epsilon}-\frac12\right]
 \,.\label{eq:CFcollq}
\end{align}
The additional factors in the second line on the rhs of Eq.~(\ref{eq:CFcollq})
(when compared to the corresponding line of Eq.~(\ref{eq:CFcollqbar})) 
are due to the fact that
\begin{equation}\label{eq:elltrasf}
 \left.\frac{}{}\bm\ell+\alpha_\ell\bm b\right|_{\mscr{\begin{array}[b]{l}
       \alpha\to\bar\alpha \\
       \bm k\to\bar{\bm k}
     \end{array}}}
 = \frac{1-\bar\alpha-\alpha_\ell}{1-\bar\alpha}
 (\bm\ell-\alpha_\ell\bm a)\,.
\end{equation}

The second difference is that, considering the soft-collinear
subtraction, we cannot use the soft limit of Eq.~(\ref{eq:CFcollq}) at
fixed $\bar\alpha$ and $\bar{\bm k}$, because this does not correspond
to the collinear limit of the soft approximation~(\ref{eq:CFsoft}).
The right way of subtracting the soft singularity
of~(\ref{eq:CFcollq}) and at the same time the $g\parallel q$
collinear singularity of~(\ref{eq:CFsoft}) is to take the collinear
limit of the latter:
\begin{align}\nonumber
 \left.\frac{d\Phi_{\gs}^{(1)}}{d\alpha\,d\bm k}
 \right|_{C_F}^{{\mathrm soft, coll,} q}
&=\I_2(\alpha, \bm k)\frac{2}{(2\pi)^{3-2\epsilon}}
 \int_0^{1-\alpha}\frac{d\alpha_\ell}{\alpha_\ell} \int d\bm\ell\,
  \frac{\Theta(\alpha_\ell\Lambda - |\ellq|)}{{\ellq}^2}
  \,.\\ \label{eq:CFsoftcollq}
&=\I_2(\alpha, \bm k)
 \frac1{(4\pi)^{2-\epsilon}\Gamma(1-\epsilon)}
 \left[\alpha^{-2\epsilon}+(1-\alpha)^{-2\epsilon}\right]
 \frac{\Lambda^{-2\epsilon}}{\epsilon^2}\,.
\end{align}

In conclusion, we write the finite part of the $C_F$ term of the
impact factor by subtracting, from $\I_3^{C_F}$, the unintegrated 
expressions of
Eqs.~(\ref{eq:CFsoft}, \ref{eq:CFcollqbar}, \ref{eq:CFsoftcollqbar}, 
\ref{eq:CFcollq}, \ref{eq:CFsoftcollq}):
\begin{align}\nonumber
 &\left.\Phi_{\gs}^{(1,{\mathrm real})}\right|_{C_F}^{\mathrm finite}\equiv
 \int d\bm k\, d\bm\ell\int_0^1 d\alpha\int_0^{1-\alpha} d\alpha_\ell
 \Biggl\{\I_3^{C_F}(\alpha_\ell, \bm\ell; \alpha, \bm k; \bm r, Q)
 \\ \nonumber
 &-\I_2(\alpha, \bm k)
 \frac{2(\bm a+\bm b)^2}{(2\pi)^{3-2\epsilon}}
 \frac{\alpha_\ell}
  {\left(\bm\ell-\alpha_\ell\bm a\right)^2 
    \left(\bm\ell+\alpha_\ell\bm b\right)^2} \\ \nonumber
&- \I_2(\alpha, \bm k) 
 \frac1{(2\pi)^{3-2\epsilon}}\left[\frac1{1-\alpha}
 P_{gq}\left(\frac{\alpha_\ell}{1-\alpha}, \epsilon \right)
 -\frac2{\alpha_\ell}\right]
 \frac{\Theta(\alpha_\ell\Lambda - |\bm\ell+\alpha_\ell\bm b|)}{
  (\bm\ell+\alpha_\ell\bm b)^2}\\ \nonumber
&-
 \left[\I_2(\alpha+\alpha_\ell, \bm k+\bm\ell) 
 \frac{\alpha+\alpha_\ell}{\alpha^2}
 P_{gq}\left(\frac{\alpha_\ell}{\alpha+\alpha_\ell}, \epsilon \right)
 \Theta(\alpha_\ell\Lambda-\frac{\alpha}{\alpha+\alpha_\ell}
 |\bm\ell-\alpha_\ell\bm a|)\right. \\
 &\qquad\left.-\I_2(\alpha,\bm k) \frac2{\alpha_\ell}
 \Theta(\alpha_\ell\Lambda - |\bm\ell-\alpha_\ell\bm a|)\right]
 \frac1{(2\pi)^{3-2\epsilon} (\bm\ell-\alpha_\ell\bm a)^2}
\,. \label{eq:CFfinite}
\end{align}
The divergent part of the $C_F$ term is obtained by re-adding the
subtracted pieces in their integrated form. In doing this, we identify
the $1-\bar\alpha$ and $-\bm r-\bar{\bm k}$ variables of
Eq.~(\ref{eq:CFcollq}) with the variables $\alpha$ and $\bm k$ of
Eqs.~(\ref{eq:CFcollqbar}, \ref{eq:CFsoftcollqbar},
\ref{eq:CFsoftcollq}) (and later with $\alpha$ and $\bm k$ of
Eq.~(\ref{eq:Phi1virtSing})), so that
\begin{align}\nonumber
 \left.\Phi_{\gs}^{(1,{\mathrm real})}\right|_{C_F}^{\mathrm divergent}
&=\int d\bm k\int_0^1d\alpha\,\I_2(\alpha, \bm k)
 \frac{c_\Gamma}{(4\pi)^{2-\epsilon}}\biggl\{
 \frac4{\epsilon^2}+\frac1{\epsilon}\left[6-4\log M^2\right]
 \\ \label{eq:CFdiv}
&\quad+2\left[8+\log^2 M^2-3\log\Lambda^2-3\log\alpha(1-\alpha)
 +\log^2\frac{\alpha}{1-\alpha}\right]+\O(\epsilon)\biggr\}\,.
\end{align}


The final expression for the one-loop correction to the photon impact
factor can be obtained by summing the real contributions given in
Eqs.~(\ref{eq:CAfinite}, \ref{eq:CAdiv}, \ref{eq:CFfinite}, \ref{eq:CFdiv})
to the virtual corrections $\Phi_{\gs}^{(1,{\mathrm virtual})}$ of
\cite{BGQ}.  One can check the cancellation of soft and collinear
singularities by combining the singular piece of the virtual
corrections~(\ref{eq:Phi1virtSing})
with the $\epsilon$-poles of Eqs.~(\ref{eq:CAdiv}) and
(\ref{eq:CFdiv}).  The cancellation of double and single pole is
straightforward. 

In conclusion, the final expression of the impact factor at NLO can be
obtained by summing the finite part of the virtual corrections, the
scale dependent term (\ref{eq:impactrenorm}) and the extra term
(\ref{eq:newfiniteSEterm}) as discussed in
section~\ref{sec:singular-virtual}, the finite terms of
Eqs.~(\ref{eq:CAdiv}) and (\ref{eq:CFdiv}), and the integrals in
Eqs.~(\ref{eq:CAfinite}) and (\ref{eq:CFfinite}) (evaluated at
$\epsilon=0$):
\begin{align}
  \Phi_{\gs}^{(1)} = &
  \left.\Phi_{\gs}^{(1,{\mathrm virtual})}\right|^{\mathrm finite}
  - \frac{2 \Phi_{\gs}^{(0)}}{(4\pi)^2} \left\{
    \beta_0 \ln\frac{\bm r^2}{\mu^2} + C_F\ln(\bm r^2) 
  \right\}
\nonumber\\
&+\frac{1}{(4\pi)^2} \int d {\bm k} \int_0^1 d\alpha\, \I_2(\alpha,{\bm k})
\bigg\{
  C_A \left[\ln^2\alpha(1-\alpha)s_0 -\ln^2 M^2\right]\nonumber\\
  &\qquad\qquad + 2C_F \left[8 - 3\ln \alpha(1-\alpha) \Lambda^2 + \ln^2 M^2
  + \ln^2 \frac{\alpha}{1-\alpha} \right]
\bigg\}\nonumber\\
&+C_A \left.\Phi_{\gs}^{(1,{\mathrm real})}\right|_{C_A}^{\mathrm finite}
+C_F \left.\Phi_{\gs}^{(1,{\mathrm real})}\right|_{C_F}^{\mathrm finite}\;.
\end{align}

\section{Conclusions}
\label{sec:conclusions}

\noindent
In this third part of our program of calculating 
the NLO corrections to the photon impact factor we first have 
presented a list of the real corrections, the $q\bar{q}g$ intermediate
state inside the impact factor. This completes the results 
of the paper \cite{BGK} which contained the longitudinal photon only.  
We then have identified and computed the divergent parts of the 
phase space integral of the real corrections: in addition to the infrared 
singularities which are due to collinear and soft configurations of the 
produced gluon there is a logarithmic divergence related to the 
$\ln s$-piece of the central region. Removal of the latter part of the 
real corrections introduces the energy scale $s_0$ which represents a central 
element of the NLO corrections. Finally, we have combined the real 
corrections with the divergent parts of the virtual corrections:
these infrared finite combinations are the main results of this paper.

With these calculations we have essentially completed the analytic
part of our program. What remains are numerical steps: the phase space
integrals both for the $q\bar{q}$ state (which includes the finite
pieces of the virtual corrections) and the $q\bar{q}g$ state. Certain
parts of these integrals can be done analytically, but, as a general
strategy, we define standard integrals and express our integrands in
terms of these expressions. The final evaluations have to be done
with the computer.

Once this is solved technically, the photon impact factor allows us to
compute the total cross section for $\gs\gs$-scattering to
next-to-leading logarithmic accuracy in $s$.  This will be the first
consistent calculation of a cross section to this order since the NLO
corrections to the BFKL kernel were first published
\cite{FL-NLO,CC-NLO}.  Taking advantage of the recently obtained
impact factor for forward jets \cite{BCV} in the same limit we
may apply the photon impact factor also to the calculation of the forward
jet cross section at HERA.  

We have emphasized already before, that interest in the NLO
calculation of the photon impact factor comes from various directions.
A prominent example is the photon wave function picture. In our
analysis \cite{BGK} of the real corrections we have extracted the new
$q\bar{q}g$ Fock component of the photon wave function which, most
conveniently, is expressed in configuration space.  In the present
paper we have studied the cancellation of divergences, but this was
done on momentum space. It will therefore be necessary to translate
these calculations into configuration space.  Furthermore, in order to
obtain a clearer geometrical picture of how color charge is
distributed inside the virtual photon, we need to take a closer look
into the various pieces found in \cite{BGK} and in this paper. Another
question of interest of the photon wave function picture is the form
of the NLO correction to the $q\bar{q}$ Fock component and of the
$q\bar{q}$ dipole cross section: do they destroy some of the main
features of the LO calculations, for example the conservation of the
transverse dipole size during the interaction with the target? Answers
to these questions are to be found in our virtual corrections, but
they require further theoretical efforts.

\acknowledgments
\noindent Helpful conversations with C.-F.~Qiao are gratefully
acknowledged.  S.G.\ wishes to thank the II.\ Institut f\"ur
Theoretische Physik of the Universit\"at Hamburg where large parts of
this work were done for kind hospitality.


\begin{thebibliography}{99}
  
\bibitem{BGQ}
J.~Bartels, S.~Gieseke and C.~F.~Qiao,
Phys.\ Rev.\ D {\bf 63} (2001) 056014
[Erratum-ibid.\ D {\bf 65} (2001) 079902]
[hep-ph/0009102].

\bibitem{BGK}
J.~Bartels, S.~Gieseke and A.~Kyrieleis,
Phys.\ Rev.\ D {\bf 65} (2002) 014006
[hep-ph/0107152].

\bibitem{BFKL}
E.~A.~Kuraev, L.~N.~Lipatov and V.~S.~Fadin,
Sov.\ Phys.\ JETP {\bf 45} (1977) 199
[Zh.\ Eksp.\ Teor.\ Fiz.\  {\bf 72} (1977) 377]; 
I.~I.~Balitsky and L.~N.~Lipatov, 
Sov.\ J.\ Nucl.\ Phys.\  {\bf 28} (1978) 822
[Yad.\ Fiz.\  {\bf 28} (1978) 1597].

\bibitem{FM} 
V.~S.~Fadin and A.~D.~Martin,
Phys.\ Rev.\ D {\bf 60} (1999) 114008
[hep-ph/9904505].

\bibitem{CSdiv}
S.~Catani and M.~H.~Seymour,
Nucl.\ Phys.\ B {\bf 485} (1997) 291
[Erratum-ibid.\ B {\bf 510} (1997) 503]
[hep-ph/9605323].






%
\bibitem{CiCo-if}
  M.\ Ciafaloni, \plb{429}{1998}{363} [hep-ph/9801322];
  M.\ Ciafaloni and D.\ Colferai, \npb{538}{1999}{187} [hep-ph/9806350].

\bibitem{Fadin-gif}
V.~S.~Fadin, R.~Fiore, M.~I.~Kotsky and A.~Papa,
Phys.\ Rev.\ D {\bf 61} (2000) 094005
[hep-ph/9908264];
Phys.\ Rev.\ D {\bf 61} (2000) 094006
[hep-ph/9908265].




\bibitem{FL-NLO}
V.~S.~Fadin and L.~N.~Lipatov,
Phys.\ Lett.\ B {\bf 429} (1998) 127
[hep-ph/9802290].

\bibitem{CC-NLO}
G.~Camici and M.~Ciafaloni,
Phys.\ Lett.\ B {\bf 412} (1997) 396
[Erratum-ibid.\ B {\bf 417} (1998) 390]
[hep-ph/9707390].

\bibitem{BCV}
J.~Bartels, D.~Colferai and G.~P.~Vacca,
Eur.\ Phys.\ J.\ C {\bf 24} (2002) 83
[hep-ph/0112283]; 
[hep-ph/0206290].

\end{thebibliography}
\end{document}